\documentclass[aps,pra,twocolumn,groupedaddress]{revtex4-1}

\usepackage{amsmath}
\usepackage{amssymb}
\usepackage{mathrsfs} 
\usepackage{graphicx} 
\usepackage{color} 

\renewcommand{\Re}{\mathrm{Re}}
\renewcommand{\Im}{\mathrm{Im}}

\begin{document}


\title{Speed-of-light limitations in passive linear media}


\author{Aaron Welters}
\email{awelters@math.mit.edu}
\author{Yehuda Avniel}
\author{Steven G. Johnson}
\affiliation{Department of Mathematics, Massachusetts Institute of Technology, Cambridge, MA 02139, USA}


\date{\today}

\begin{abstract}
We prove that well-known speed of light restrictions on electromagnetic energy velocity can be extended to a new level of generality, encompassing even nonlocal chiral media in periodic geometries, while at the same time weakening the underlying assumptions to only passivity and linearity of the medium (either with a transparency window or with dissipation). As was also shown by other authors under more limiting assumptions, passivity alone is sufficient to guarantee causality and positivity of the energy density (with no thermodynamic assumptions). Our proof is general enough to include a very broad range of material properties, including anisotropy, bianisotropy (chirality), nonlocality, dispersion, periodicity, and even delta functions or similar generalized functions. We also show that the ``dynamical energy density'' used by some previous authors in dissipative media reduces to the standard Brillouin formula for dispersive energy density in a transparency window. The results in this paper are proved by exploiting deep results from linear-response theory, harmonic analysis, and functional analysis that had previously not been brought together in the context of electrodynamics.
\end{abstract}


\maketitle

\section{Introduction}
Because relativity implies that information should travel more slowly than $c$ (the speed of light in vacuum), there is a long history of interest in proving how this subluminal restriction arises for electromagnetic energy velocity ${\bf v}_e$, and various authors have proved that the speed-of-light limitation $||{\bf v}_e|| \le c$ holds for homogeneous causal dispersive media \cite{Brillouin1932, Brill60, Loudon65, Loud70, SchulzDuBois69, Oughstun88}, \cite[\S 84]{Land84}, for homogeneous passive isotropic or anisotropic dispersive media \cite{Glas01, Yagh07a, Yagh07b}, or for periodic nondispersive media \cite[p.~41]{John08}, usually assuming negligible loss (i.e., in a ``transparency window'').  For lossy media various authors have proved similar bounds, but with some disagreement over the definition of energy velocity \cite{Brill60, Loud70, SchulzDuBois69, Oughstun88, Geppert65, Glas01}. In this paper, we extend those results by both greatly simplifying the underlying assumptions and by generalizing the reach of the results. Similar in spirit to \cite{Yagh07a, Yagh07b, Glas01} (which were limited to homogeneous media), we prove $||{\bf v}_e|| \le c$ assuming only that we have a \emph{passive} medium (one which does no net work on the fields) with or without a transparency window, and from this we obtain not only $||{\bf v}_e|| \le c$ but also causality of the polarization response and positivity of the energy density (which some other authors justified on thermodynamic grounds \cite[\S 80]{Land84}).  Furthermore, our results apply to arbitrary anisotropic and \emph{bianisotropic} (chiral) dispersive media, both homogeneous (Sec.~\ref{Sec:HomogMedia}, similar to Yaghjian \cite{Yagh07a, Yagh07b} and Glasgow \textit{et al.}~\cite{Glas01}) and \emph{periodic} (Sec.~\ref{Sec:PeriodicMedia}) media, include certain classes of spatially \emph{nonlocal} media and only require that the medium be passive ``on average'' in the unit cell of periodicity.  Mathematically, we generalize the concept of a linear susceptibility to arbitrary distributions over the vector-valued Hilbert space of electromagnetic fields in the unit cell, and extend the notion of a transparency window to a very general measure-theoretic definition that we connect to traditional formulations of electromagnetic dissipation. The key elements of our proof (whose technical details are in Appendix \ref{AppendixSecProofs}) are derived from important previous theorems on passive convolution operators and Herglotz functions from linear-response theory and functional analysis \cite{Zema65, Zema63, Zema70, Zema72, Aron56, Gree60, Gesz00, Gesz01}.  The energy velocity ${\bf v}_e$ in a transparency window is equivalent to the well-known group velocity ${\bf v}_g$ \cite{Havelock14, Brill60, Land84, Jack75, John08, Strat07}, and in Appendix \ref{AppendixEnergyVelEqsGroupVel} we extend this well-known equivalence ${\bf v}_e={\bf v}_g$ \cite{Abraham11, Havelock14, Brill60, Jack75, Land84, Yaghjian09, Yeh79, John08, Biot57, Wagner59I, Wagner59II, Bertoni66, McPhedran10, ChuTamir71}. Furthermore, we strengthen the $||{\bf v}_e|| \le c$ bound by showing that $||{\bf v}_e|| = c$ can only hold for a very special class of materials and field solutions (Sec.~\ref{Sec:Achieving_c}).  In Sec.~\ref{Sec:Velocity_In_Lossy_Media} we extend these bounds to lossy media following Glasgow \textit{et al.}~\cite{Glas01}, and we show that the ``dynamical energy density'' used for lossy media in previous work \cite{Glas01, Geppert65, Yaghjian05, Yagh07a, Figotin07a, Figotin07b} reduces to the familiar dispersive energy density \cite{Abraham11, Havelock14, Brill60, Land84} in the limit of a transparency window (greatly simplifying and generalizing earlier proofs \cite{Yaghjian05, Yagh07a, Figotin07a, Figotin07b}). Finally, in Sec.~\ref{Sec:Conclusion} we discuss several open problems.

Several important results were derived in previous work on energy-velocity limitations in electromagnetism. Brillouin \cite{Brillouin1932, Brill60} assumed a homogeneous, isotropic, linear medium with a Lorentzian model of material dispersion. The original derivation of energy density for the single-resonance Lorentz medium by Brillouin, however, included an algebraic error. This was first pointed out by Loudon \cite{Loudon65, Loud70} who, after correcting this error, obtained a subluminal expression for the energy velocity in a Lorentz medium, and Loudon's formulation was subsequently extended by other authors \cite{SchulzDuBois69, Oughstun88}. However, some later works argued that one should include an additional term in the energy density (which appears in the denominator of the energy velocity), equal to a total dissipated energy \cite{Glas01, Yagh07a, Figotin07a, Figotin07b}, and this also results in a subluminal energy velocity \cite{Glas01}; our bounds apply regardless of whether this term is included.  Whereas the early work was limited to a specific (usually Lorentz ``oscillator'') model of dispersion, the newer ``dynamical'' energy density has the advantage of being model-independent.  In the case of negligible loss (where there is no disagreement about the energy density), the energy velocities derived by Brillouin and Loudon are shown by Loudon to be equivalent and bounded by $c$. Landau and Lifshitz \cite{Land84} assumed homogeneous, isotropic, linear dispersive materials satisfying causality, positivity of energy density due to thermodynamic considerations, and some regularity conditions on the permittivity and permeability as a function of frequency. Their derivation of $||{\bf v}_e||\le c$ in a transparency window is then based on the Kramers-Kronig relations and the positivity of the energy density. Yaghjian \cite{Yagh07a} was able to show positivity of the energy density (with some later corrections \cite{Yagh07b}) and $||{\bf v}_e||\le c$ assuming only passivity of the medium in a transparency window, anticipating our results from Sec.~\ref{Sec:HomogMedia}. Although he derived a positive energy density for arbitrary bianisotropic homogeneous media, Yaghjian \cite{Yagh07a} only applied this to a speed-of-light bound in the isotropic case; also, that derivation did not explicitly specify the allowed functional forms of the susceptibilities, nor did it connect those results to causality or to the theorems of linear-response theory (some of which are rederived by \cite{Yagh07a} in limited forms). Nevertheless, the results of Yaghjian can viewed as a precursor of our results.  For lossy media, Glasgow~\textit{et al.}~\cite{Glas01} employed similar passivity assumptions in order to obtain positivity of the dissipated energy and hence subluminal energy velocity.  All of these previous results, however, were for energy velocity defined pointwise in space, or equivalently for light propagation in a homogeneous medium (or possibly a homogenized approximation to an inhomogeneous ``metamaterial''), but a more general setting is that of \emph{periodic} media in which Bloch waves propagate with a well-defined group and energy velocity \cite{Brillouin1946, Yeh79, John08, Biot57, Wagner59I, Wagner59II, Bertoni66, McPhedran10, ChuTamir71}.  Joannopoulos \textit{et al.}~\cite{John08} proved $||{\bf v}_e||\le c$ in this periodic setting, but only under the assumptions of isotropic, dispersionless, linear media with real permittivity and permeability bounded below by one.  In this paper, we combine the weaker assumption of passivity with a very general setting of periodic media in order to unify and generalize these past results.

\section{Homogeneous Media}\label{Sec:HomogMedia}
In this section we prove the speed-of-light limitation in the simplified case of homogeneous passive linear media, in preparation for the more difficult case of periodic media in Sec.~\ref{Sec:PeriodicMedia}. In Sec.~\ref{Sec:PlanewaveEnergyVel}, we review the definitions of planewave, group velocity, and energy velocity. In Sec.~\ref{Sec:HomogPassivity}, we set up the basic properties of susceptibility and define passivity. In Sec.~\ref{Sec:HomogTranspWnd}, we give a definition of a transparency window. In Sec.~\ref{Sec:HomogSpeedLimit}, we derive the speed-of-light limitation in a transparency window.

\subsection{Planewaves and Energy Velocity}\label{Sec:PlanewaveEnergyVel}
In a homogeneous medium, because of the continuous translational symmetry, Maxwell's equations admit nontrivial solutions which are time-harmonic electromagnetic planewaves
\begin{eqnarray}\label{DefEMPlaneWave}
{\bf E}e^{i({\bf k}\cdot{\bf r}-\omega t)},\;\;
{\bf H}e^{i({\bf k}\cdot{\bf r}-\omega t)},
\end{eqnarray}
where ${\bf E}$, ${\bf H}$ are constant vectors, with real frequency $\omega$ and real wavevector $\bf k$ (in bandwidth of negligible loss), satisfying a dispersion relation $\omega=\omega({\bf k})$. The group velocity is defined as ${\bf v}_g=\nabla_{\bf k} \omega(\bf k)$, where $\nabla_{\bf k}$ denotes the gradient, which is the velocity of narrow-bandwidth wavepackets \cite{Havelock14, Brill60}. The energy velocity ${\bf v}_e$ of such a field is defined \cite{Havelock14, Brill60, Land84, Jack75, John08, Strat07} as the ratio of the time-averaged energy flux to the time-averaged energy density $U$, i.e.,
\begin{eqnarray}\label{DefEnergyVelocity}
{\bf v}_e=\frac{\Re\,{\bf S}}{U},\;\;\text{where }{\bf S}=\frac{c}{8\pi}{\bf E}\times {\bf H}^\ast
\end{eqnarray}
is the complex Poynting vector in Gaussian units. It is known \cite{Havelock14, Brill60, Land84, Jack75, John08, Strat07} that ${\bf v}_g={\bf v}_e$ for transparent media and we review the fact that this is true even for dispersive media in Appendix \ref{AppendixEnergyVelEqsGroupVel}. In dispersionless isotropic or anisotropic media, the energy density is $U=\frac{1}{16\pi}\left({\bf E}^{\dagger}\epsilon {\bf E}+{\bf H}^{\dagger} \mu {\bf H}\right)$, where $\epsilon, \mu$ are the frequency-independent permittivity and permeability tensors, respectively, but when the media are dispersive with negligible loss then the energy density is given by the Brillouin formula \cite{Abraham11, Havelock14, Brill60, Land84}, namely, $U=\frac{1}{16\pi}\left({\bf E}^{\dagger} \frac{d\omega\epsilon(\omega)}{d\omega} {\bf E}+{\bf H}^{\dagger} \frac{d\omega\mu(\omega)}{d\omega} {\bf H}\right)$. 

More generally, we consider anisotropic, bianisotropic, and dispersive linear media, which are described by a $6\times 6$ susceptibility $\chi$: 
\begin{eqnarray}\label{DefSusceptibConvol}
\left[\begin{array}{c}
    {\bf P} \\
    {\bf M}
  \end{array}\right]
=
\chi\ast\left[\begin{array}{c}
    {\bf E} \\
    {\bf H}
  \end{array}\right]
=\int_{-\infty}^{\infty}\chi(t^\prime)\left[\begin{array}{c}
    {\bf E}(t-t^\prime) \\
    {\bf H}(t-t^\prime)
  \end{array}\right] dt^\prime,
\end{eqnarray}
for the electric and magnetic polarizations ${\bf P}=\frac{1}{4\pi}\left({\bf D}-{\bf E}\right)$ and ${\bf M}=\frac{1}{4\pi}\left({\bf B}-{\bf H}\right)$. Mathematically, the convolution is defined in the distribution sense as described in Appendix \ref{AppdxSecSusceptibilityAndPassiveLinearMedia}, allowing the susceptibility to be a generalized function (such as a delta function), i.e., a distribution on the smooth vector-valued functions $\begin{bmatrix} {\bf E}(t) & {\bf H}(t)\end{bmatrix}^{\text{T}}$ with compact support. For such general materials, when the frequency $\omega$ belongs to a tranparency window (as defined in Sec.~\ref{Sec:HomogTranspWnd}), the Brillouin formula for the energy density $U$ can be generalized \cite{Figotin07a, Figotin07b, Yaghjian05, Yagh07a, Yagh07b, Yaghjian09} in terms of the Fourier transform of electromagnetic susceptibility $\widehat{\chi}(\omega)$ (a $6\times 6$ matrix-valued function of frequency $\omega$) by
\begin{eqnarray}\label{DefEnergyDensHomog}
U=\frac{1}{16\pi}\left[\begin{array}{c}
    {\bf E} \\
    {\bf H}
  \end{array}\right]^\dagger\frac{d}{d\omega}\omega\left[ I+4\pi\widehat{\chi}(\omega)\right]\left[\begin{array}{c}
    {\bf E} \\
    {\bf H}
  \end{array}\right],
\end{eqnarray}
where $I$ is the identity matrix.

\subsection{Passivity}\label{Sec:HomogPassivity}
The key property of the susceptibility that will be used to derive the energy velocity bound $||{\bf v}_e||\leq c$ and other properties is \emph{passivity}, which is the statement that \emph{polarization currents don't do work}. Mathematically, this is the following condition: the inequality
\begin{eqnarray}\label{DefPassiveHomog}
0\leq \Re \int_{-\infty}^t {\bf E}(t^\prime)^\dagger\frac{d{\bf P}(t^\prime)}{d t^\prime}+{\bf H}(t^\prime)^\dagger\frac{d{\bf M}(t^\prime)}{d t^\prime}\, dt^\prime,
\end{eqnarray}
must hold for all time $t$ and all smooth vector-valued functions $\begin{bmatrix} {\bf E}(t) & {\bf H}(t)\end{bmatrix}^{\text{T}}$ with compact support. Physically, the integral in (\ref{DefPassiveHomog}) represents the work that the fields do on the bound currents \cite{Jack75}. Thus a passive medium is one satisfying the passivity condition (\ref{DefPassiveHomog}) or, in other words, is one in which energy can only be absorbed by the material but never generated by it.

A key point here, which is known in a closely related distributional setting of Zemanian \cite{Zema72} and proved in this setting in Appendix \ref{AppdxSecSusceptibilityAndPassiveLinearMedia}, is that the convolution equation (\ref{DefSusceptibConvol}) and the passivity condition (\ref{DefPassiveHomog}) mean that $\frac{d\chi}{dt}\ast$ is a special kind of convolution operator known as a ``passive'' convolution operator \cite[Def.~8.2-2]{Zema72}, a fact that has many important consequences. We will now describe three of these consequences which are proved in Appendix \ref{AppdxSecProofsofSecHomogPassivityAndSecPeriodicPassivity}: causality, analyticity, and positivity. As we shall see later in this paper, it is the latter two properties along with a transparency window that are key to deriving positivity of the energy density, which is used to derive the speed-of-light limitation $||{\bf v}_e||\leq c$. 

First, passivity implies causality---the polarization only depends on fields in the past, i.e., $\chi\ast$ is a causal convolution operator \cite[Def.~4.6-1]{Zema72} or, equivalently, $\chi(t)=0$ for $t<0$ (in the distributional sense \cite[\S 3.3, p.~55]{Zema72}). Technically, passivity only implies that $\frac{d\chi}{dt}\ast$ is a causal convolution operator, but this means that since $\chi$ is an antiderivative of $\frac{d\chi}{dt}$ then there exists some antiderivative $X$ that differs from $\chi$ by a constant such that $X\ast$ is a causal convolution operator. Our proof below doesn't require that additive constant to be zero, but on physical grounds it of course must be (as a nonzero constant would mean the polarizations would depend on the field on all times in the future and past).

Second, the Fourier transform $\widehat{\chi}(\omega)=(\mathscr{F}\chi)(\omega)$ exists for $\Im\, \omega>0$ and is an analytic $6\times 6$ matrix-valued function of the complex frequency $\omega$ in the upper half-plane. Third, the matrix
\begin{eqnarray}\label{DefHerglotzFuncHomog}
h(\omega)=\omega\widehat{\chi}(\omega),\;\Im\,\omega>0
\end{eqnarray}
has a positive semidefinite imaginary part, i.e.,
\begin{eqnarray}\label{HerglotzFuncHomogImSemidef}
\Im\, h(\omega)\geq 0,\;\Im\,\omega>0,
\end{eqnarray}
where the imaginary part is defined as $\Im\, h(\omega)=\frac{1}{2i}[h(\omega)-h(\omega)^\dagger]$. The above two properties mean that $h(\omega)$ is what is known as a matrix-valued Herglotz function \cite{Gesz00}, i.e., a matrix-valued function analytic in the upper half-plane with positive semidefinite imaginary part in that domain. In fact, a necessary and sufficient condition for $h(\omega)=\omega\widehat{\chi}(\omega)$ to be a Herglotz function is that $\frac{d\chi}{dt}\ast$ is a passive convolution operator, i.e, that the passivity condition (\ref{DefPassiveHomog}) holds. Herglotz functions have been heavily studied  (see, for instance, \cite{Aron56}, \cite{Gesz00}, \cite{Gesz01}, \cite{Gesz13}, and references within) and their properties, along with a transparency window, are the key to the positivity of the energy density and the speed-of-light limitation.

\subsection{Transparency Window}\label{Sec:HomogTranspWnd}
A transparency window for a passive linear media is defined as a frequency interval $(\omega_1,\omega_2)\subseteq \mathbb{R}$ where losses are negligible \cite[\S 80, \S 84]{Land84}. A precise definition in terms of the Herglotz function $h(\omega)$ will be given in this section. Recall that we require a transparency window for a well-defined energy velocity and for the group velocity to equal the energy velocity. In typical circumstances, the condition of a transparency window is simply that $\Im\, h(\omega)=0$ in the interval, but more generally the definition is given in terms of a measure induced by $\Im\, h(\omega)$ which we discuss briefly in this section. Two important properties of the function $h(\omega)$ in the transparency window will also be discussed. In particular, we show that the Herglotz function $h(\omega)$ is a monotonically increasing and differentiable Hermitian-matrix-valued function of frequency in the window. These properties are critical since they imply the positivity of the energy density, which is used to derive the speed-of-light limitation on the energy velocity in the next section.

\begin{figure}[h]
\includegraphics[height=3in,width=3in]{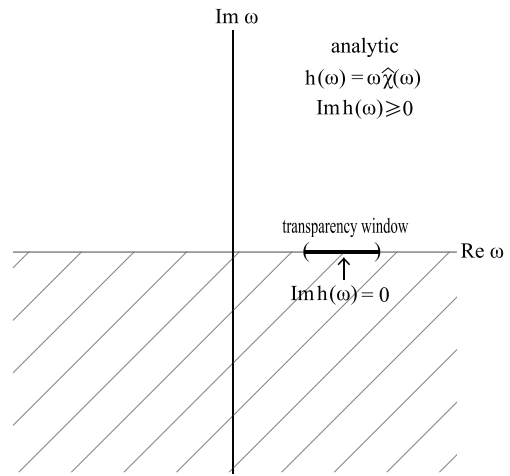}
\caption{\label{fig:TransparencyWindow} The analytic properties of the function $h(\omega)=\omega\widehat{\chi}(\omega)$ along with a transparency window. This is all that is needed to prove the speed-of-light limitation $||{\bf v}_e||\leq c$. Moreover, the analytic properties of $h(\omega)$ are necessary and sufficient conditions for $\frac{d\chi}{dt}\ast$ to be a passive convolution operator.}
\end{figure}

Now, as is to be expected, there is a correspondence between the imaginary part $\Im\, h(\omega)$ of the Herglotz function and electromagnetic losses (i.e., dissipation of energy) due to the transfer of electromagnetic energy into matter by absorption. The simplest case in which to understand this correspondence is that of isotropic media with integrable susceptibility $\chi(t)$ and time-harmonic fields ${\bf E}(t)={\bf E}e^{-i\omega t},{\bf H}(t)={\bf H}e^{-i\omega t}$ (with spatial dependence suppressed). In this case, the permittivity $\epsilon(\omega)$ and permeability $\mu(\omega)$ are scalar functions which are related to the susceptibility by $\widehat{\chi}(\omega)=\begin{bmatrix} \frac{\epsilon(\omega)-1}{4\pi}I & 0\\ 0 & \frac{\mu(\omega)-1}{4\pi}I\end{bmatrix}$. It follows from (\ref{DefSusceptibConvol}) that the polarization ${\bf P}(t)={\bf P}e^{-i\omega t}$ and magnetization ${\bf M}(t)={\bf M}e^{-i\omega t}$ are time-harmonic with ${\bf P}=\frac{\epsilon(\omega)-1}{4\pi}{\bf E}$ and ${\bf M}=\frac{\mu(\omega)-1}{4\pi}{\bf H}$. Now the value of the integrand in (\ref{DefPassiveHomog}) for the real parts of these time-harmonic fields represents the power dissipated per unit time and so its time-average is just
\begin{align}\label{Eg1IsotrLosses}
&\frac{1}{p}\int_0^p \Re\,{\bf E}(t^\prime)^\dagger\frac{d\Re\,{\bf P}(t^\prime)}{dt^\prime}+\Re\, {\bf H}(t^\prime)^\dagger\frac{d\Re\,{\bf M}(t^\prime)}{dt^\prime}\, dt^\prime \nonumber\\
&=\frac{1}{2}\Re\,\left[{\bf E}^\dagger (-i\omega {\bf P})+{\bf H}^\dagger(-i\omega {\bf M})\right]\nonumber\\
&=\frac{1}{2}\left[\begin{array}{c}
    {\bf E} \\
    {\bf H}
  \end{array}\right]^\dagger \Im\, h(\omega)\left[\begin{array}{c}
    {\bf E} \\
    {\bf H}
  \end{array}\right] \nonumber\\
&=\frac{\omega}{8\pi}[\Im\, \epsilon(\omega)||{\bf E}||^2+\Im\, \mu(\omega)||{\bf H}||^2],
\end{align} 
where $p=2\pi/\omega$. But (\ref{Eg1IsotrLosses}) is just the well-known expression for the electric and magnetic losses in a dispersive isotropic medium for monochromatic electromagnetic fields \cite[\S 80, Eq.~(80.4)]{Land84}.

The next simplest case to understand is that of isotropic media with non-monochromatic fields in which the electromagnetic fields ${\bf E}(t),{\bf H}(t)$ are square-integrable and both $\chi(t),\frac{d\chi}{dt}(t)$ are integrable. Then, by Young's convolution theorem \cite[Theorem 23.44]{Yeh06} and Plancherel's theorem \cite[Theorem IX.6]{Sim81}, the convolution in (\ref{DefSusceptibConvol}) can be Fourier transformed and the total energy dissipated by such fields is the quantity (\ref{DefPassiveHomog}) with $t=\infty$ which is just the value
\begin{align}\label{Eg2IsotrLosses}
&\Re \int_{-\infty}^\infty {\bf E}(t^\prime)^\dagger\frac{d{\bf P}(t^\prime)}{d t^\prime}+{\bf H}(t^\prime)^\dagger\frac{d{\bf M}(t^\prime)}{d t^\prime}\, dt^\prime \nonumber\\
&=\Re\int_{-\infty}^\infty \widehat{\bf E}(\omega)^\dagger(-i\omega\widehat{\bf P}(\omega))+\widehat{\bf H}(\omega)^\dagger(-i\omega\widehat{\bf M}(\omega))\, \frac{d\omega}{2\pi} \nonumber\\
&=\int_{-\infty}^\infty\left[\begin{array}{c}
    \widehat{\bf E}(\omega) \\
    \widehat{\bf H}(\omega)
  \end{array}\right]^\dagger \Im\,h(\omega) \left[\begin{array}{c}
    \widehat{\bf E}(\omega) \\
    \widehat{\bf H}(\omega)
  \end{array}\right] \frac{d\omega}{2\pi} \nonumber\\
&=\frac{1}{4\pi}\int_{-\infty}^\infty \omega\left[\Im\,\epsilon(\omega)||\widehat{\bf E}(\omega)||^2+\Im\,\mu(\omega)||\widehat{\bf H}(\omega)||^2\right]\, \frac{d\omega}{2\pi}.
\end{align} 
But (\ref{Eg2IsotrLosses}) is just the well-known expression for the electric and magnetic losses in a dispersive isotropic medium for non-monochromatic electromagnetic fields which go sufficiently rapidly to zero as $t\rightarrow \pm \infty$ \cite[\S 80, Eq.~(80.6)]{Land84}.

In the most general case of passive linear media, the susceptibility $\chi(t)$ is a generalized function and so it may not be integrable. Thus the correspondence between the imaginary part $\Im\, h(\omega)$ and losses is more complicated, involving concepts of measures which we discuss in Appendix \ref{AppdxSecProofsofSecHomogTranspWndSecPeriodicTranspWndAndSecAchievingc} in more detail. In particular, it is shown in 
Appendix \ref{AppdxSecProofsofSecHomogTranspWndSecPeriodicTranspWndAndSecAchievingc} that there exists a nonnegative-matrix-valued measure $\Omega$ on the bounded Borel subsets of $\mathbb{R}$ such that (\ref{DefPassiveHomog}), with $t=\infty$ and for any smooth vector-valued function $F(t)=\begin{bmatrix} {\bf E}(t) & {\bf H}(t)\end{bmatrix}^{\text{T}}$ with compact support, is given in terms of an integral involving only this measure and the Fourier transform of $F$ integrated over all real frequencies, namely,
\begin{align}
&\Re \int_{-\infty}^\infty {\bf E}(t^\prime)^\dagger\frac{d{\bf P}(t^\prime)}{d t^\prime}+{\bf H}(t^\prime)^\dagger\frac{d{\bf M}(t^\prime)}{d t^\prime}\, dt^\prime\label{HomoImhGenConn2Passivity0}\\
&=\frac{1}{2}\int_\mathbb{R}d{\widehat F}(\omega)^\dagger \Omega_\omega{\widehat F}(\omega)\label{HomoImhGenConn2Passivity}.
\end{align} 
Moreover, we prove in Appendix \ref{AppdxSecProofsofSecHomogTranspWndSecPeriodicTranspWndAndSecAchievingc} that the value of this measure $\Omega$ on any finite open interval is given below by (\ref{HomogMeasureLosses}) which again comes from the imaginary part $\Im\,h(\omega)$ by taking a limit as we approach the real axis from the upper-half plane.
 
From the discussion above we can conclude that a way to quantify how lossy a material is in a certain finite frequency range $(\omega_1,\omega_2)\subseteq \mathbb{R}$ is to use the limits
\begin{eqnarray}\label{HomogMeasureLosses}
\Omega((\omega_1,\omega_2))=\lim_{\delta\downarrow 0}\lim_{\eta\downarrow 0} \int_{\omega_1+\delta}^{\omega_2-\delta}\frac{1}{\pi}\Im\, h(\omega+i\eta)\, d\omega.
\end{eqnarray}
The fact that these limits always exist, which we prove in Appendix \ref{AppdxSecProofsofSecHomogTranspWndSecPeriodicTranspWndAndSecAchievingc},  is just based on fundamental properties of matrix-valued Herglotz functions described in \cite{Gesz00}. The reason that it is necessary to define the limits in (\ref{HomogMeasureLosses}) as such is to handle the most general possible cases for $h$ that can arise from the theory of Herglotz functions \cite{Gesz00}. One can see how this measures losses in the frequency range $(\omega_1,\omega_2)$ for isotropic media under the conditions on $\chi(t)$ which lead to the identities (\ref{Eg1IsotrLosses}) and (\ref{Eg2IsotrLosses}) since in this case the equation (\ref{HomogMeasureLosses}) simplifies to:
\begin{align}
\Omega((\omega_1,\omega_2))&=\lim_{\delta\downarrow 0}\lim_{\eta\downarrow 0} \int_{\omega_1+\delta}^{\omega_2-\delta}\frac{1}{\pi}\Im\, h(\omega+i\eta)\, d\omega\\
&=\lim_{\delta\downarrow 0}\int_{\omega_1+\delta}^{\omega_2-\delta}\frac{1}{\pi}\lim_{\eta\downarrow 0} \Im\, h(\omega+i\eta)\, d\omega \nonumber\\
&=\int_{\omega_1}^{\omega_2}\frac{1}{\pi} \Im\, h(\omega)\, d\omega \nonumber.
\end{align}

Now we give the precise definition of a transparency window using (\ref{HomogMeasureLosses}). The interval $(\omega_1,\omega_2)$ is said to be a transparency window for the medium if losses are negligible in this frequency range, i.e., if
\begin{eqnarray}\label{DefTranspWindHomog}
\Omega((\omega_1,\omega_2))=0.
\end{eqnarray}
As shown in Appendix \ref{AppdxSecProofsofSecHomogTranspWndSecPeriodicTranspWndAndSecAchievingc}, it follows from this definition that if $(\omega_1,\omega_2)$ is a transparency window then $h(\omega)$ can be analytically continued into the lower half-plane through this interval $(\omega_1,\omega_2)$ to a function satisfying the Hermitian and monotonicity conditions
\begin{eqnarray}\label{HerglotzHomogMonot}
\Im\,h(\omega)=0,\;\;\frac{dh}{d\omega}(\omega)\geq 0,\;\;\omega\in(\omega_1,\omega_2).
\end{eqnarray}

As mentioned above these properties, summarized in Fig.~\ref{fig:TransparencyWindow}, are key since, as we shall see in the next section, they imply the positivity of the energy density and the speed-of-light limitation on the energy velocity.

\subsection{Speed-of-light Limitation}\label{Sec:HomogSpeedLimit}
We now show that the speed-of-light limitation
\begin{eqnarray}\label{SpeedLightLimitaHomog}
||{\bf v}_e||\leq c
\end{eqnarray}
holds for any time-harmonic electromagnetic planewave with frequency $\omega$ in a transparency window $(\omega_1,\omega_2)$. To do this we break the proof into two steps. First, we bound the numerator $\Re\, {\bf S}$ in equation (\ref{DefEnergyVelocity}) using elementary inequalities. Second, we bound the denominator $U$ using more sophisticated inequalities that follow from the properties of the Herglotz function $h(\omega)$.

To bound the numerator in (\ref{DefEnergyVelocity}) we first use the Lagrange's identity for complex numbers \cite[p.~30, Prob.~1.48]{Apos74} to conclude that
\begin{eqnarray}\label{LagrangesIdComplex}
||{\bf E}\times{\bf H}^\ast||^2=||{\bf E}||^2||{\bf H}||^2-|{\bf E}^{\text{T}} {\bf H}|^2
\end{eqnarray}
From this identity and the elementary inequality $||{\bf E}||||{\bf H}||\leq \frac{1}{2}(||{\bf E}||^2+||{\bf H}||^2)$, we obtain the following upper bound on the energy flux
\begin{eqnarray}\label{EnergyFluxHomogUppBdd}
||\Re\,{\bf S}|| \leq \frac{c}{16\pi}\left(||{\bf E}||^2+||{\bf H}||^2\right).
\end{eqnarray}

Next, we deal with the denominator in (\ref{DefEnergyVelocity}). It follows from the monotonicity property in (\ref{HerglotzHomogMonot}) that the energy density $U$ in (\ref{DefEnergyDensHomog}) is well-defined for any electromagnetic planewave with frequency $\omega$ in the transparency window $(\omega_1,\omega_2)$ and satisfies 
\begin{align}\label{EnergyDensHomogLowBdd}
U&=\frac{1}{16\pi}\left[\begin{array}{c}
    {\bf E} \\
    {\bf H}
  \end{array}\right]^\dagger\frac{d}{d\omega}\omega\left[ I+4\pi\widehat{\chi}(\omega)\right]\left[\begin{array}{c}
    {\bf E} \\
    {\bf H}
  \end{array}\right]\\
&=\frac{1}{16\pi}\left(||{\bf E}||^2+||{\bf H}||^2\right)+\frac{1}{4}\left[\begin{array}{c}
    {\bf E} \\
    {\bf H}
\end{array}\right]^\dagger\frac{dh}{d\omega}(\omega)\left[\begin{array}{c}
    {\bf E} \\
    {\bf H}
  \end{array}\right] \nonumber \\
&\geq\frac{1}{16\pi}\left(||{\bf E}||^2+||{\bf H}||^2\right). \nonumber
\end{align}
In particular, from this inequality it follows that the energy density is positive and so the energy velocity ${\bf v}_e$ defined by (\ref{DefEnergyVelocity}) doesn't blow up.
Finally, from the two bounds (\ref{EnergyFluxHomogUppBdd}), (\ref{EnergyDensHomogLowBdd}) we deduce the upper bound on the energy velocity
\begin{eqnarray}\label{EnergyVelHomogUppBdd}
||{\bf v}_e||=\frac{||\Re\,{\bf S}||}{U}\leq \frac{\frac{c}{16\pi}\left(||{\bf E}||^2+||{\bf H}||^2\right)}{\frac{1}{16\pi}\left(||{\bf E}||^2+||{\bf H}||^2\right)}= c.
\end{eqnarray}

\section{Periodic Media}\label{Sec:PeriodicMedia}
In this section we prove the speed-of-light limitation in periodic passive linear media. This section is a strict generalization of Sec.~\ref{Sec:HomogMedia} and as such the presentation here will follow the same outline. First, we review Bloch waves, group velocity, and energy velocity. Next, we set up the basic properties of susceptibility, define passivity, and then give a definition of a transparency window. Finally, we derive the speed-of-light limitation in a transparency window.

The main difference between this section and the special case, Sec.~\ref{Sec:HomogMedia}, is that planewaves must be replaced with Bloch waves, and, as such, the Fourier transform of the susceptibility in general no longer takes on values that are matrices, but instead its values are bounded linear operators on an infinite-dimensional Hilbert space (where, for generality, we include the possibility of spatially nonlocal operators). Nevertheless, the basic outline is the same even if the theorems are a bit more complicated. 

\subsection{Bloch Waves and Energy Velocity}
In a periodic medium, because of the discrete translational symmetry, Maxwell's equations admit nontrivial solutions which are time-harmonic electromagnetic Bloch waves \cite{John08, Kuchment01}
\begin{eqnarray}\label{DefEMBlochWave}
{\bf E}({\bf r})e^{i({\bf k}\cdot{\bf r}-\omega t)},\;\;
{\bf H}({\bf r})e^{i({\bf k}\cdot{\bf r}-\omega t)},
\end{eqnarray}
where ${\bf E}({\bf r})$, ${\bf H}({\bf r})$ are periodic functions on the lattice, with real frequency $\omega$ and real wavevector $\bf k$ (in a transparency window), satisfying a dispersion relation $\omega=\omega({\bf k})$. The group velocity is ${\bf v}_g=\nabla_{\bf k} \omega(\bf k)$ as before. In Sec.~\ref{Sec:PlanewaveEnergyVel}, the energy velocity was defined as the ratio of the energy flux to the energy density at a single point in space since all points were the same for a planewave, but now we must average over a unit cell, i.e., the energy velocity ${\bf v}_e$ of such a field is \cite{John08, Yeh79}
\begin{align}\label{DefEnergyVelocityPeriodicMedia}
{\bf v}_e=\frac{\int_V\Re\,{\bf S}({\bf r})d{\bf r}}{\int_V U({\bf r})d{\bf r}},
\end{align}
where ${\bf S},U$ are defined as in Sec.~\ref{Sec:PlanewaveEnergyVel} (but now are functions of position ${\bf r}$), $V$ denotes a unit cell of the periodic medium, and the vector-valued functions ${\bf E}({\bf r})$, ${\bf H}({\bf r})$ are assumed to be square-integrable in $V$. Just as in Sec.~\ref{Sec:PlanewaveEnergyVel}, the equality ${\bf v}_g={\bf v}_e$ holds for transparent media, under certain reservations, a fact that we review in Appendix \ref{AppendixEnergyVelEqsGroupVel} [see (\ref{IdentityGrpVelEqsEnergyVelGenerally})]. As discussed in Appendix A, V can be unbounded in certain directions, for example in a periodic waveguide, in which case the energy velocity is only defined in the periodic directions.

Now we can handle any type of periodic linear media, including isotropic, anisotropic, bianisotropic, and dispersive media that are described by the same convolution form as (\ref{DefSusceptibConvol}), but with a spatial dependence on ${\bf r}$. As before, the convolution is defined in the distribution sense (as described in Appendix \ref{AppdxSecSusceptibilityAndPassiveLinearMedia}) except that now the susceptibility $\chi(t)$ is a distribution on a different set of test functions, namely, the smooth functions of time with compact support that take values in the Hilbert space of square-integrable vector-valued functions in the unit cell $V$. For such general materials, when the frequency $\omega$ belongs to a transparency window (as defined in Sec.~\ref{Sec:PeriodicTranspWnd}), the energy density $U$ is more complicated and is given in Appendix \ref{AppendixEnergyVelEqsGroupVel} by equation (\ref{DefEnergyDensityGeneralCase}). This is a generalization of the Brillouin formula (\ref{DefEnergyDensHomog}) to arbitrary periodic passive linear media for frequencies in a transparency window.

The reason for the complication in the definition of the energy density $U$ is that the Fourier transform of electromagnetic susceptibility $\widehat{\chi}(\omega)$ will be a function of frequency whose values are bounded linear operators in this infinite-dimensional Hilbert space. That just means the polarizations at a point ${\bf r}$ can be a function of $\begin{bmatrix} {\bf E}({\bf r}^\prime) & {\bf H}({\bf r}^\prime)\end{bmatrix}^{\text{T}}$ at other points in the unit cell. This generalization allows us to include responses nonlocal within the unit cell. As nonlocal responses are often used to describe metals at short lengthscales \cite{McMahon09, Abajo10, Abajo11, Ruppin75, BenPark12, Yannopapas08, Stefanou08, LiSunChan06}, this generalization could be used to describe, for instance, a periodic array of disjoint metallic objects. However, the most common case is that of local media, in which the polarizations at a point ${\bf r}$ only depend upon the field at the point ${\bf r}$. This means the susceptibility in the frequency domain is typically just matrix multiplication by $\widehat{\chi}({\bf r},\omega)$, a $6\times 6$ matrix-valued function of frequency $\omega$ and space ${\bf r}$ which is spatially periodic. In this case, the energy density is essentially identical to the Brillouin formula (\ref{DefEnergyDensHomog}) except that the derivative $\frac{d}{d\omega}$ is replaced by the partial derivative $\frac{\partial}{\partial\omega}$ (since now there is a dependence on ${\bf r}$).

\subsection{Passivity}\label{Sec:PeriodicPassivity}
In a periodic medium, just as in the homogeneous case, the key property of the susceptibility that will be used to derive the energy velocity bound $||{\bf v}_e||\leq c$ and other properties is \emph{passivity}. But it turns out that we don't need to require passivity at every point individually like in (\ref{DefPassiveHomog}), and instead we can require a weaker condition: that the medium is passive on average over the unit cell, i.e., \emph{polarization currents don't do work on average within the unit cell $V$}. Mathematically this means our condition is like that of (\ref{DefPassiveHomog}), but with an integral over the unit cell as well as over time. More precisely, this is the following condition: the inequality
\begin{align}\label{DefPassiveInhomog}
0\leq \Re \int_{-\infty}^t \int_V{\bf E}(t^\prime)^\dagger\frac{d{\bf P}(t^\prime)}{dt^\prime}+{\bf H}(t^\prime)^\dagger\frac{d{\bf M}(t^\prime)}{d t^\prime}\, d{\bf r} dt^\prime,
\end{align}
(suppressing the spatial dependency ${\bf r}$) holds for all time $t$ and all smooth functions of time $\begin{bmatrix} {\bf E}(t) & {\bf H}(t)\end{bmatrix}^{\text{T}}$ with compact support which take values in the Hilbert space of square-integrable vector-valued functions in the unit cell $V$.

Again the key point here is that the convolution equation~(\ref{DefSusceptibConvol}) and the passivity condition (\ref{DefPassiveInhomog}) mean that $\frac{d\chi}{dt}\ast$ is a ``passive'' convolution operator. The consequences are essentially the same as those described in Sec.~\ref{Sec:HomogMedia} (and proved in Appendix \ref{AppdxSecProofsOfMainResults}), with the only difference being in interpretation and proof, which is just due to the fact that the Hilbert space we are now working in is infinite-dimensional. For instance, just like before (\ref{DefPassiveInhomog}) implies causality, i.e., $\chi(t)=0$ for $t<0$, with the same technical caveat as previously mentioned. Also, its Fourier transform $\widehat{\chi}(\omega)=(\mathscr{F}\chi)(\omega)$ exists for $\Im\, \omega>0$, but now it is an analytic bounded-operator-valued function. Similarly $h(\omega)$, defined as in (\ref{DefHerglotzFuncHomog}), is a bounded-operator-valued Herglotz function, meaning it is analytic (in the operator norm topology) in the upper half plane with (\ref{HerglotzFuncHomogImSemidef}) satisfied, i.e., the imaginary part of the operator is positive semidefinite (in the operator sense rather than the matrix sense) with respect to the inner product (\ref{AppdxDefHilbertSpaceInnerProduct}). And, like before, a necessary and sufficient condition for $h(\omega)=\omega\widehat{\chi}(\omega)$ to be a Herglotz function is that $\frac{d\chi}{dt}\ast$ is a passive convolution operator, i.e, the passivity condition (\ref{DefPassiveInhomog}) holds.

\subsection{Transparency Window}\label{Sec:PeriodicTranspWnd}
A transparency window for a periodic passive linear media is still defined as a frequency interval $(\omega_1,\omega_2)\subseteq \mathbb{R}$ where losses are negligible \cite[\S 80, \S 84]{Land84}, but now electromagnetic losses are due to the transfer of electromagnetic energy into matter by absorption in the unit cell $V$. And a precise definition in terms of the imaginary part $\Im\, h(\omega)$ of the Herglotz function $h(\omega)$ is the same as in Sec.~\ref{Sec:HomogTranspWnd}, i.e., (\ref{DefTranspWindHomog}) must be satisfied, the only difference is that the limits in (\ref{HomogMeasureLosses}), which still exist as proven in Appendix \ref{AppdxSecProofsofSecHomogTranspWndSecPeriodicTranspWndAndSecAchievingc}, are now in the strong operator topology \cite[\S VI.1]{Sim81} of the Hilbert space with inner product (\ref{AppdxDefHilbertSpaceInnerProduct}). And similar to before, the two important properties of the function $h(\omega)$ in a transparency window $(\omega_1,\omega_2)$ still continue to hold in this case, namely,  $h(\omega)$ can be analytically continued (in operator norm topology) into the lower half-plane through this interval $(\omega_1,\omega_2)$ to a bounded-operator-valued function satisfying the self-adjoint and monotonicity conditions in (\ref{HerglotzHomogMonot}), where now analyticity, self-adjointness, and positive semidefiniteness is with respect to Hilbert space with inner product (\ref{AppdxDefHilbertSpaceInnerProduct}). These properties of the function $h(\omega)$, along with the transparency window, all of which do not differ in form from Sec.~\ref{Sec:HomogMedia}, are depicted in Fig.~\ref{fig:TransparencyWindow}.

\subsection{Speed-of-light Limitation}
We now will show that speed-of-light limitation
\begin{eqnarray}\label{SpeedLightLimitaPeriodic}
||{\bf v}_e||\leq c
\end{eqnarray}
holds for any time-harmonic electromagnetic Bloch wave provided the periodic medium is passive in its unit cell $V$ and the frequency $\omega$ is in a transparency window $(\omega_1,\omega_2)$. Deriving the speed-of-light limitation is done in essentially the same way as was done in Sec.~\ref{Sec:HomogSpeedLimit}. 

First, we derive the upper bound on the numerator in (\ref{DefEnergyVelocityPeriodicMedia}) in exactly same way as (\ref{EnergyFluxHomogUppBdd}) with one additional step, namely, we use the elementary inequality
\begin{align}\label{EnergyFluxPeriodicUppBdd}
\left\Vert\int_V\Re\,{\bf S({\bf r})}d{\bf r}\right\Vert \leq \int_V\left\Vert\Re\,{\bf S({\bf r})}\right\Vert d{\bf r}. 
\end{align}

Second, deriving the lower bound on the denominator in (\ref{DefEnergyVelocityPeriodicMedia}) is done essentially in the same way as (\ref{EnergyDensHomogLowBdd}), except that now we must use the fact that $\frac{dh}{d\omega}(\omega)\geq 0$ is in the operator sense so that
\begin{align}\label{EnergyDensPeriodicLowBdd}
\int_V U({\bf r})d{\bf r}\geq\int_V\frac{1}{16\pi}\left(||{\bf E}({\bf r})||^2+||{\bf H}({\bf r})||^2\right)d{\bf r}.
\end{align}
Again, this inequality means that the energy density is positive so that the energy velocity ${\bf v}_e$ given by (\ref{DefEnergyVelocityPeriodicMedia}) doesn't blow up.
Finally, from the two bounds (\ref{EnergyFluxPeriodicUppBdd}), (\ref{EnergyDensPeriodicLowBdd}) we deduce the upper bound on the energy velocity
\begin{eqnarray}\label{EnergyVelPeriodicUppBdd}
||{\bf v}_e||=\frac{\left\Vert\int_V\Re\,{\bf S({\bf r})}d{\bf r}\right\Vert}{\int_V U({\bf r})d{\bf r}}\leq c.
\end{eqnarray}

\section{Achieving $c$}\label{Sec:Achieving_c}
Here we analyze the speed-of-light limitation $||{\bf v}_e||\le c$ for necessary conditions on the fields and the media for the upper bound $c$ to be achieved in a transparency window. As we shall show, $||{\bf v}_e||=c$ is never achieved except in vacuum or possibly in a few very special cases which we discuss.

We can see from the inequalities (\ref{LagrangesIdComplex})--(\ref{EnergyVelHomogUppBdd}) and (\ref{EnergyFluxPeriodicUppBdd})--(\ref{EnergyVelPeriodicUppBdd}) that the upper bound~$||{\bf v}_e||=c$ is achieved by a time-harmonic electromagnetic planewave~(\ref{DefEMPlaneWave}) or Bloch wave~(\ref{DefEMBlochWave}) exactly when the numerator and denominator in~(\ref{EnergyVelHomogUppBdd}) or (\ref{EnergyVelPeriodicUppBdd}), respectively, achieve simultaneously the upper and lower bound of the inequalities (\ref{EnergyFluxHomogUppBdd}) and (\ref{EnergyDensHomogLowBdd}), respectively, in the case of a planewave or (\ref{EnergyFluxPeriodicUppBdd}), (\ref{EnergyFluxHomogUppBdd}) and (\ref{EnergyDensPeriodicLowBdd}), respectively, in the case of a Bloch wave. First, we consider under what circumstances the numerator achieves its upper bound, and we see a certain relationship between $\epsilon$ and $\mu$ (via the electric and magnetic parts of $\widehat{\chi}$) is required that is difficult to achieve outside of vacuum.  Second, we examine when the denominator achieves its lower bound, and it turns out that this is only possible for a very specific form of the dispersion relation $\widehat{\chi}(\omega)$ which is proportional to $1/\omega$.

Necessary conditions for the numerator to achieve the upper bound are that the complex amplitude vector $F=\begin{bmatrix} {\bf E} & {\bf H}\end{bmatrix}^{\text{T}}$ of the corresponding planewave or Bloch wave satisfy the following three conditions:
\begin{eqnarray}\label{CondHomogEnergyFluxUppBdd}
{\bf E}^{\text{T}} {\bf H}=0,\;\;\; ||{\bf E}||=||{\bf H}||,\;\;\; \Im\, ({\bf E}\times {\bf H}^\ast)=0,
\end{eqnarray}
(almost everywhere in the unit cell), which follows from the identity~(\ref{LagrangesIdComplex}) and the inequality~(\ref{EnergyFluxHomogUppBdd}). In the case of planewaves, the conditions~(\ref{CondHomogEnergyFluxUppBdd}) are both necessary and sufficient for the numerator in (\ref{EnergyVelHomogUppBdd}) to achieve the upper bound in (\ref{EnergyFluxHomogUppBdd}). The conditions (\ref{CondHomogEnergyFluxUppBdd}) are very restrictive: for instance, for planewave solutions (\ref{DefEMPlaneWave}) at a nonzero real wavevector ${\bf k}$ in a homogeneous isotropic medium, these conditions imply that the relative permittivity and permeability are equal. More complicated possibilities, including non-transverse solutions to (\ref{CondHomogEnergyFluxUppBdd}), arise in anisotropic media. For example, one such solution is:
\begin{gather}
\omega_0=\frac{c}{3}\sqrt{2},\;\;\;{\bf k}=-\sqrt{2}{\bf e}_3,\;\;\; {\bf E}=\sqrt{2}{\bf e}_2,\;\;\; {\bf H}={\bf e}_1+{\bf e}_3\\
\epsilon(\omega_0)=\begin{bmatrix} 1 & 0 & 0\\ 0 & 3 & 0 \\ 0 & 0 & 1 \end{bmatrix},\;\;\;\mu(\omega_0)=\begin{bmatrix} 5 & 0 & -2\\ 0 & 1 & 0 \\ -2 & 0 & 2 \end{bmatrix},
\end{gather}
where ${\bf e}_j$, $j=1,2,3$ denote the standard orthonormal basis vectors in $\mathbb{R}^3$.

We derive another constraint on the materials by considering (\ref{EnergyDensHomogLowBdd}) or (\ref{EnergyDensPeriodicLowBdd}) from the denominator in~(\ref{EnergyVelHomogUppBdd}) or (\ref{EnergyVelPeriodicUppBdd}), respectively. The lower bound in the inequalities (\ref{EnergyDensHomogLowBdd}) or (\ref{EnergyDensPeriodicLowBdd}), respectively, is achieved exactly when the complex amplitude vector $F=\begin{bmatrix} {\bf E} & {\bf H}\end{bmatrix}^{\text{T}}$ of the wave in (\ref{DefEMPlaneWave}) or (\ref{DefEMBlochWave}), respectively, with frequency $\omega$ is in the nullspace of the operator $h^\prime(\omega)$, i.e.,
\begin{eqnarray}\label{HerglotzPlanewaveNullspace}
h^\prime(\omega)F=0,
\end{eqnarray}
which follows from the monotonicity property~(\ref{HerglotzHomogMonot}) and the definition of the energy density (\ref{DefEnergyDensHomog}) for homogeneous media or (\ref{DefEnergyDensityGeneralCase}) for periodic media in a transparency window. But from a deep result in the theory of Herglotz functions (which we describe and prove in Appendix \ref{AppdxSecProofsofSecHomogTranspWndSecPeriodicTranspWndAndSecAchievingc}) it follows that the equality~(\ref{HerglotzPlanewaveNullspace}) being true for a single frequency in a transparency window implies that~(\ref{HerglotzPlanewaveNullspace}) is true for \emph{all} $\omega$, real or complex, so that there exists a constant operator $h_0$ such that 
\begin{eqnarray}\label{HerglotzPlanewaveNullspace1}
h(\omega)F=h_0F,\;\;\;\Im\,h_0=0
\end{eqnarray}
is satisfied for all $\omega$. In terms of the Fourier transform of the susceptibility $\widehat\chi(\omega)$ we therefore must have
\begin{eqnarray}\label{HerglotzPlanewaveNullspace2}
\widehat\chi(\omega)F=\frac{1}{\omega}h_0F,
\end{eqnarray}
for \emph{all} $\omega\not=0$.

Equation (\ref{HerglotzPlanewaveNullspace2}) is a very strong, and rather unrealistic, constraint upon the material dispersion (especially since the susceptibility is usually argued to decay faster than $1/\omega$ as $\omega\to\infty$, on physical grounds \cite[\S 7.5]{Jack75}, \cite[\S 7.10 (c)]{Jack75}, \cite[\S 78]{Land84}, or to prove the Kramers-Kronig relations by contour integration \cite[\S 7.10 (d)]{Jack75}, \cite[\S 82]{Land84}, \cite[\S 123]{Land96}). It includes vacuum, of course ($h_0 = 0$), and can also be viewed as including perfect conductors as a limiting case.  For example, a perfect electric conductor (PEC), which imposes a Dirichlet condition of vanishing tangential electric field, can be obtained by taking the limit $s \to \infty$ of an isotropic electric susceptibility $\frac{\epsilon-1}{4\pi} = \frac{s}{\omega}$ (see \cite{Sihv05}, \cite{Sihv07}, \cite{Sihv08}).  Inhomogeneous geometries consisting only of vacuum and PEC are known to support modes with $||{\bf v}_e||=c$, such as the TEM mode of a coaxial PEC waveguide \cite[\S 9.20]{Strat07}. 

Equation (\ref{HerglotzPlanewaveNullspace2}) superficially resembles the susceptibility $i\sigma/\omega$ of a material with a constant scalar conductivity $\sigma$ (see \cite[\S 7.5 (c)]{Jack75}), but this case is excluded by the condition of a transparency window, which requires that $h_0$ be self-adjoint, i.e., $\Im\, h_0=0$.  The coefficient $h_0$ can be further constrained if one imposes the usual condition that real fields produce real polarizations in the time domain \cite[\S 7.10 (c)]{Jack75}, \cite[\S 82]{Land84}, \cite[\S 123]{Land96} which, in the case of homogeneous media, implies that $h_0$ must be of the form $h_0 = i\sigma$ where $\sigma$ is a real anti-symmetric matrix. For isotropic media, this implies that $h_0=0$, which is the vacuum case.  For a nonchiral medium, it would mean that a nonzero $h_0$ would necessarily break electromagnetic reciprocity \cite{Lore1896}, \cite{Kong70}, \cite[\S 89]{Land84}, \cite{Pott04}, \cite{Chew08} (due to $\widehat{\chi}(\omega)\not=\widehat{\chi}(\omega)^{\text{T}}$). For a chiral medium, its possible to have such an $h_0=i\sigma$ which is actually consistent with reciprocity \cite{Chew08}. Thus, a nonzero $h_0$ would mean: if the medium is nonchiral then it is nonreciprocal (having the form of a gyrotropic medium \cite[\S 79, Prob.~1]{Land84}, \cite[\S 101]{Land84}) or it can be reciprocal, but then it is chiral.

\section{Velocity in Lossy Media}\label{Sec:Velocity_In_Lossy_Media}

In the previous sections, we focused on the case of materials with negligible losses (specifically, a transparency window) which corresponds to the situation in which group velocity is a meaningful energy velocity, but of course the problem of energy transport velocity in lossy media is also important.  In lossy media, Poynting's theorem can be generalized to define a ``dynamical'' energy density \cite{Glas01, Geppert65, Yaghjian05, Yagh07a, Figotin07a, Figotin07b}, from which one can define a ``dynamical'' energy velocity via the ratio of the Poynting flux to the energy density \cite{Glas01, Geppert65}. Recently, Glasgow \textit{et al.}~\cite{Glas01} showed that this dynamical energy velocity is $\leq c$ in passive media, showed that it bounds the front velocity \cite{Brill60}, \cite[\S 5.18]{Strat07} (the speed at which a region of nonzero fields expands) generalizing a result of Sommerfeld and Brillouin \cite{Sommerfeld1907, Sommerfeld1914, Brillouin1914, Brill60} by using a similar derivation used in \cite[\S 2.4.3]{Evan10} for the wave equation, and also bounded the velocity of the mean energy-weighted position. 
In this section, we generalize and extend some of these results in several ways. 

We begin in Sec.~\ref{Sec:Dynamical_Energy_Density_And_Velocity} by reviewing the dynamical energy density and velocity constructions, and we delineate the types of susceptibilities for which the inequalities hold---unlike previous authors, we include distributions and furthermore generalize to the case of periodic and nonlocal media (as in Sec.~\ref{Sec:PeriodicMedia}). 

It is essential that the dynamical energy density reduce to the familiar dispersive energy density (\ref{DefEnergyDensHomog}) in a transparency window, and for passive linear media this equivalence was shown by Yaghjian and Best \cite{Yaghjian05}, Yaghjian \cite{Yagh07a}, and Figotin and Schenker \cite{Figotin07a, Figotin07b}. The proof in \cite{Yaghjian05, Yagh07a}, which applies to certain bianisotropic homogeneous media, doesn't include periodic and nonlocal media nor does it specify the regularity of susceptibilities for which their argument applies. The proof of equivalence in \cite{Figotin07a, Figotin07b} is rather cumbersome, requiring several pages and an extension of Maxwell's equations in lossy media to a conservative system by introducing auxiliary fields.  It was also limited to square-integrable fields over all space (excluding Bloch waves, for example), and required more restrictive assumptions about the susceptibility (excluding distributions, for example). Therefore, in Sec.~\ref{Sec:Connection_To_Dispersive_Energy} we provide an alternative proof, which is much shorter and (we believe) more intuitive, does not require the conservative extension, and applies to the full generality of fields and susceptibilities considered in this paper.

\subsection{Dynamical Energy Density and Velocity}\label{Sec:Dynamical_Energy_Density_And_Velocity}

We begin by reviewing the generalization of Poynting's theorem to arbitrary lossy media, from which the dynamical energy density is defined as in \cite{Glas01}, but we do so in the context of our generalized susceptibility definitions of Secs.~\ref{Sec:HomogMedia} and \ref{Sec:PeriodicMedia}.

Consider the macroscopic Maxwell's equations (in Gaussian units):
\begin{align}
\nabla\cdot {\bf D}&=4\pi\rho & &\nabla \times {\bf H}=\frac{4\pi}{c}{\bf J}+\frac{1}{c}\frac{\partial {\bf D}}{\partial t}\label{MaxwellEqLine1}\\
\nabla\cdot {\bf B}&=0 & &\nabla \times {\bf E}+\frac{1}{c}\frac{\partial {\bf B}}{\partial t}=0.\label{MaxwellEqLine2}
\end{align}
We are using complex fields which are not necessarily time-harmonic so we want to derive Poynting's theorem \cite[\S 6.8]{Jack75}, i.e., the energy conservation law, for such fields in which the rate of doing work per unit volume by the fields is taken to be
\begin{eqnarray}\label{Def:RateWorkPerUnitVolume}
\Re\left({\bf J}^\dagger {\bf E}\right).
\end{eqnarray}
In order to exhibit this conservation law explicitly we use Maxwell's equations to express (\ref{Def:RateWorkPerUnitVolume}) in other terms using the method from \cite[\S 6.8]{Jack75}. First, we begin by taking the complex conjugate of both sides of the equality in (\ref{MaxwellEqLine1}) and take the dot product with ${\bf E}$ to yield:
\begin{eqnarray}\label{DerivationPoyntingThm1}
{\bf J}^\dagger {\bf E}=\frac{c}{4\pi}{\bf E}^{\text{T}}\left(\nabla \times {\bf H}^\ast\right)-\frac{1}{4\pi}{\bf E}^{\text{T}}\left(\frac{\partial {\bf D}}{\partial t}\right)^\ast.
\end{eqnarray}
Next, we use the vector identity,
\begin{eqnarray}
\nabla\cdot\left(a\times b\right)=b^{\text{T}}\left(\nabla\times a\right)-a^{\text{T}}\left(\nabla\times b\right) 
\end{eqnarray}
with $a={\bf E}$, $b={\bf H}^\ast$ and use (\ref{MaxwellEqLine2}) to get that the right side of (\ref{DerivationPoyntingThm1}) is
\begin{eqnarray}\label{DerivationPoyntingThm2}
{\bf J}^\dagger {\bf E}=-\nabla\cdot{\bf S}-\frac{1}{4\pi}{\bf H}^\dagger\frac{\partial {\bf B}}{\partial t}-\frac{1}{4\pi}{\bf E}^{\text{T}}\left(\frac{\partial {\bf D}}{\partial t}\right)^\ast,
\end{eqnarray}
where ${\bf S}=\frac{c}{4\pi} {\bf E}\times {\bf H}^\ast$.
Finally, we take the real part of both sides of (\ref{DerivationPoyntingThm2}) to get
\begin{align}\label{PoyntingsTheoremDifferentialForm}
\Re\left({\bf J}^\dagger {\bf E}\right)=-\nabla\cdot\Re\,{\bf S}-\frac{1}{4\pi}\Re\left({\bf H}^\dagger\frac{\partial {\bf B}}{\partial t}+{\bf E}^\dagger\frac{\partial {\bf D}}{\partial t}\right),
\end{align}
where the term $U(t)$ (suppressing spatial dependency) in
\begin{align}\label{Def:DynamicalEnergyDensityDiffForm}
\frac{\partial U}{\partial t} = \frac{1}{4\pi}\Re\left({\bf H}^\dagger\frac{\partial {\bf B}}{\partial t}+{\bf E}^\dagger\frac{\partial {\bf D}}{\partial t}\right)
\end{align}
is interpreted to be a \emph{dynamical energy density}. Notice that this result (\ref{PoyntingsTheoremDifferentialForm}) is the standard form for the Poynting's theorem in differential form if we use real fields and assume the usual condition that real fields produce real polarizations in the time domain.

Now we can express this as an energy conservation law in an integral form by
\begin{align}
-\frac{\partial u}{\partial t}=\int_{V}\Re \left({\bf J}^\dagger {\bf E}\right)d{\bf r}+\int_{V}\nabla\cdot\Re\,{\bf S}\,d{\bf r},\label{EnergyConservLawIntegralForm}
\end{align}
where $V$ is the periodic unit cell,
\begin{align}\label{Def:DynamicalEnergyDensity}
u(t)&=\int_{-\infty}^t\int_{V} \frac{1}{4\pi}\Re\left({\bf H}^\dagger\frac{\partial {\bf B}}{\partial t^\prime}+{\bf E}^\dagger\frac{\partial {\bf D}}{\partial t^\prime}\right) d{\bf r} dt^\prime\\
&=\frac{1}{8\pi}\left(||{\bf E}(t)||^2+||{\bf H}(t)||^2\right)\nonumber\\
&+\Re\,\int_{-\infty}^t\int_{V} {\bf E}^\dagger\frac{\partial {\bf P}}{\partial t^\prime}+{\bf H}^\dagger\frac{\partial {\bf M}}{\partial t^\prime}\, d{\bf r} dt^\prime\nonumber,
\end{align}
assuming sufficiently smooth fields which go to zero sufficiently rapidly as $t\rightarrow \pm\infty$ in the norm $||\cdot||$ with respect to the inner product (\ref{Def:HilbertSpaceInnerProduct}) (for homogeneous media we just drop the integral over $V$ in the definitions). From the passivity property (\ref{DefPassiveInhomog}) it follows that the total \emph{dynamical energy} $u$ (in $V$) defined above is always a nonnegative quantity and satisfies the inequality
\begin{eqnarray}\label{EnergyDensDynamicLowBdd}
u(t)\ge \frac{1}{8\pi}\left(||{\bf E}(t)||^2+||{\bf H}(t)||^2\right)
\end{eqnarray}
for all $t\in \mathbb{R}$.

We define the \emph{dynamical energy velocity} to be
\begin{eqnarray}\label{Def:EnergyVelocityDynamic}
{\bf v}_e(t) = \frac{\int_V \Re\,{\bf S}(t)\,d{\bf r}}{u(t)},
\end{eqnarray}
whenever $u(t)\not= 0$ for $t\in \mathbb{R}$, where
\begin{eqnarray}
{\bf S}(t)=\frac{c}{4\pi} {\bf E}(t)\times {{\bf H}(t)}^\ast.
\end{eqnarray}

As we will now show, the speed-of-light limitation
\begin{eqnarray}
||{\bf v}_e(t)||\le c
\end{eqnarray}
holds for every $t\in \mathbb{R}$ in which $u(t)\not=0$.

To prove this inequality, we first derive the upper on the numerator in (\ref{Def:EnergyVelocityDynamic}) similar to the inequality (\ref{EnergyFluxHomogUppBdd}) by using the inequalities (\ref{LagrangesIdComplex}) and (\ref{EnergyFluxPeriodicUppBdd}). Second, the lower bound (\ref{EnergyDensDynamicLowBdd}) as already been proven for the denominator in (\ref{Def:EnergyVelocityDynamic}). Finally, from these bounds we derive the upper bounded on the dynamical energy velocity
\begin{eqnarray}\label{EnergyVelUppBddLossyMedia}
||{\bf v}_e(t)||= \frac{\left \Vert \int_V \Re\,{\bf S}(t)\,d{\bf r}\right \Vert}{u(t)}\le c
\end{eqnarray}
for every $t\in \mathbb{R}$ in which $u(t)\not=0$.

\subsection{Connection to Dispersive Energy Density and Velocity in a Transparency Window}\label{Sec:Connection_To_Dispersive_Energy}

In this section we will show that the dynamical energy (\ref{Def:DynamicalEnergyDensity}) and velocity (\ref{Def:EnergyVelocityDynamic}) reduces to the dispersive energy and velocity for a transparency window.  We prove this result by adapting an approach from \cite{Zema70, Zema72}: we construct a complex-frequency (exponentially growing) waveform that smoothly ``turns off'' after some time $T$ (in order to have a well-defined polarization response), and then we take a limit as the frequency approaches a transparency window. Combined with the analyticity and self-adjointness of $h(\omega)$, we obtain the familiar Brillouin formula (involving $dh/d\omega$) for the dispersive energy density (\ref{DefEnergyDensHomog}) and the energy velocity (\ref{DefEnergyVelocity}) for homogeneous media and their generalizations (\ref{DefEnergyDensityGeneralCase}), (\ref{DefEnergyVelocityPeriodicMedia}) for periodic media.

Choose any $T\in \mathbb{R}$ and any $\lambda(t)\in C^{\infty}(\mathbb{R})$ satisfying
\begin{eqnarray}\label{Def:SmoothCutOffOfIdentity}
\lambda (t)=\left\{ 
\begin{array}{ll}
1, & \text{if }t<T \\ 
0, & \text{if }T+1 <t
\end{array}
\right..
\end{eqnarray}
Define $\theta(t)=e^{-i\omega t}\lambda(t)$, for $\Im\,\omega>0$ and $t\in \mathbb{R}$. Consider now the Hilbert space of all vector-valued functions which are square-integrable in $V$ with inner product $(\cdot,\cdot)$ defined in (\ref{Def:HilbertSpaceInnerProduct}) and let $F({\bf r})=\begin{bmatrix} {\bf E}({\bf r}) & {\bf H}({\bf r}) \end{bmatrix}^{\text{T}}$ be a nonzero element in this space. Then for the fields $ {\bf E}({\bf r},t)={\bf E}({\bf r})\theta(t)$, $ {\bf H}({\bf r},t)={\bf H}({\bf r})\theta(t)$ we can perform the time integral in (43), relying on the proofs of \cite[Theorem 10.1]{Zema70}, \cite[Theorem 8.12-1]{Zema72} that these integral exists. This yields
\begin{align}\label{DispersiveEnergyDensitySmoothTurnOffExpWaveform}
\Re\,\int_{-\infty}^t\int_{V} {\bf E}^\dagger\frac{\partial {\bf P}}{\partial t^\prime}+{\bf H}^\dagger\frac{\partial {\bf M}}{\partial t^\prime}\, d{\bf r} dt^\prime=\frac{e^{2\Im\,\omega t}}{2\Im\,\omega}(F,\Im\,h(\omega)F)
\end{align}
for all $t<T$. From this, the definition of these fields, the definition (\ref{Def:SmoothCutOffOfIdentity}), and the definition (\ref{Def:DynamicalEnergyDensity}) we find for these fields that
\begin{align}\label{dyn_energy_density_far_in_past_special_fields}
u(t)=\frac{e^{2\Im\,\omega t}}{8\pi}\left [(F,F)+\frac{4\pi}{\Im\,\omega}(F,\Im\,h(\omega)F)\right ]
\end{align}
for all $t<T$ and is positive (at the end of this section, we will get an additional $1/2$ factor by time-averaging).

From the formula  (\ref{dyn_energy_density_far_in_past_special_fields}) and the definition (\ref{Def:EnergyVelocityDynamic}) we find for these fields that the dynamical energy velocity is
\begin{align}
{\bf v}_e(t)&=\frac{\int_V \Re\,{\bf S}(t)\,d{\bf r}}{u(t)}\\
&=\frac{\int_V \Re\left(\frac{c}{8\pi} {\bf E}({\bf r})\times {{\bf H}({\bf r})}^\ast\right)\,d{\bf r}}{\frac{1}{16\pi}\left[(F,F)+\frac{4\pi}{\Im\,\omega}(F,\Im\,h(\omega)F)\right]}\nonumber
\end{align}
for all $t<T$. 

Now though, if $\omega_0\in(\omega_1,\omega_2)$ and $(\omega_1,\omega_2)\subseteq \mathbb{R}$ is a transparency window then it follows that for $\omega=\omega_0+i\eta$, $\eta>0$ we have
\begin{eqnarray}\label{Im_part_herglotz_func_connect_to_deriv}
\lim_{\eta\downarrow 0}\frac{1}{\Im\,\omega}\Im\,h(\omega)=h^\prime(\omega_0),
\end{eqnarray}
with the limit converging in the operator norm topology.
And hence we find that
\begin{align}
\lim_{\eta\downarrow 0}{\bf v}_e(t)&=\frac{\int_V \Re\left(\frac{c}{8\pi} {\bf E}({\bf r})\times {{\bf H}({\bf r})}^\ast\right)\,d{\bf r}}{\frac{1}{16\pi}\left[(F,F)+4\pi(F,h^\prime(\omega_0)F)\right]}\\
&=\frac{\int_V \Re\left(\frac{c}{8\pi} {\bf E}({\bf r})\times {{\bf H}({\bf r})}^\ast\right)\,d{\bf r}}{\left(F,\frac{1}{16\pi}\frac{d}{d\omega}\omega\left[ I+4\pi\widehat{\chi}(\omega)\right]\big|_{\omega=\omega_0}F\right)}
\end{align}
independent of $t$, provided $t<T$. The denominator here is in the form of the Brillouin formula for dispersive energy density. Now all that remains is to verify the consequence for Bloch waves and time-averaging.

Now suppose we had a time-harmonic Bloch wave in (\ref{DefEMBlochWave}) of the form
\begin{eqnarray}\label{dynm_energy_BlochWave_fields}
{\bf E}_{\bf k_0}({\bf r})e^{i({\bf k_0}\cdot{\bf r}-\omega_0 t)},\;\;
{\bf H}_{\bf k_0}({\bf r})e^{i({\bf k_0}\cdot{\bf r}-\omega_0 t)},
\end{eqnarray}
where ${\bf E}_{\bf k_0}({\bf r})$, ${\bf H}_{\bf k_0}({\bf r})$ are periodic functions on the lattice which are square-integrable in the periodic unit cell $V$, with real frequency $\omega_0$ in a transparency window, and real wavevector $\bf k_0$. Then for the $F$ above we can take $F({\bf r})=\begin{bmatrix} {\bf E}_{\bf k_0}({\bf r})e^{i{\bf k_0}\cdot{\bf r}} & {\bf H}_{\bf k_0}({\bf r})e^{i{\bf k_0}\cdot{\bf r}} \end{bmatrix}^{\text{T}}$. And so by the argument just given for this $F$ and $\omega_0$ we find that
\begin{align}
\lim_{\eta\downarrow 0}{\bf v}_e(t)=\frac{\int_V \Re\left(\frac{c}{8\pi} {\bf E}_{\bf k_0}({\bf r})\times {{\bf H}_{\bf k_0}({\bf r})}^\ast\right)\,d{\bf r}}{\int_V U({\bf r})\,d{\bf r}}={\bf v}_e,
\end{align}
independent of $t$, provided $t<T$, where $U$ is defined by (\ref{DefEnergyDensityGeneralCase}) and ${\bf v}_e$ is the energy velocity (\ref{DefEnergyVelocityPeriodicMedia}) of this Bloch wave.

Using a similar approach above we will now prove that the dynamical energy (\ref{Def:DynamicalEnergyDensity}) for the real parts of fields $ {\bf E}({\bf r},t)={\bf E}({\bf r})\theta(t)$, $ {\bf H}({\bf r},t)={\bf H}({\bf r})\theta(t)$ reduces after time-averaging to the dispersive energy (\ref{DefEnergyDensityGeneralCase}), (\ref{DefEnergyDensityGeneralCaseSupplement}). In spirit this is similar to the fact that for time-harmonic fields $a$ and $b$ the time average of their scalar product is $\left<{\Re\,a}^{\text{T}}\Re\,b\right>=\frac{1}{2}\Re\, (a^{\dagger}b)$, but some care is required here because of the exponential growth in time for complex $\omega$. The proof requires we impose the usual condition that real fields produce real polarizations in the time domain \cite[\S 7.10 (c)]{Jack75}, \cite[\S 82]{Land84}, \cite[\S 123]{Land96}. Let $\Phi(t)=\theta(t)F$, where as before $F({\bf r})=\begin{bmatrix} {\bf E}({\bf r}) & {\bf H}({\bf r}) \end{bmatrix}^{\text{T}}$. Then its polarization response is $\begin{bmatrix} {\bf P}(t) &  {\bf M}(t) \end{bmatrix}^{\text{T}}=\chi \ast \Phi(t)$ and the latter condition implies that
\begin{align}
\begin{bmatrix} \Re\, {\bf P}(t) & \Re\, {\bf M}(t) \end{bmatrix}^{\text{T}}=\Re\, [\chi \ast \Phi(t)]=\chi \ast \Re\, \Phi(t).
\end{align}
This implies the dynamical energy (\ref{Def:DynamicalEnergyDensity}) for $\Re\, \Phi$, which we will denote by $u_{\text{real}}$, is given by
\begin{align}
u_{\text{real}}(t)&=\frac{1}{8\pi}\left(||\Re\,{\bf E}(t)||^2+||\Re\,{\bf H}(t)||^2\right)\nonumber\\
&+\int_{-\infty}^t\int_{V} \Re\, {\bf E}^{\text{T}}\Re\,\frac{\partial {\bf P}}{\partial t^\prime}+\Re\,{\bf H}^{\text{T}}\Re\,\frac{\partial {\bf M}}{\partial t^\prime}\, d{\bf r} dt^\prime.
\end{align}
Next, we use the vector identity
\begin{align}
\Re\,a^{\text{T}}\Re\,b=\frac{1}{2}\Re\,(a^\dagger b) + \frac{1}{2}\Re\,(a^{\text{T}} b)
\end{align}
to write $u_{\text{real}}$ in terms of the dynamical energy $u$ of $F$,
\begin{align}\label{dynm_energy_den_real_fields}
u_{\text{real}}(t) &= \frac{1}{2}u(t)+ \frac{1}{16\pi}\Re\,\int_{V} {\bf E}(t)^{\text{T}}{\bf E}(t) + {\bf H}(t)^{\text{T}}{\bf H}(t) d{\bf r} \nonumber\\
&+\frac{1}{2}\Re\,\int_{-\infty}^t\int_{V} {\bf E}^{\text{T}}\frac{\partial {\bf P}}{\partial t^\prime}+{\bf H}^{\text{T}}\frac{\partial {\bf M}}{\partial t^\prime}\, d{\bf r} dt^\prime.
\end{align}
Then similar to the proofs of \cite[Theorem 10.1]{Zema70}, \cite[Theorem 8.12-1]{Zema72} it can be easily shown that
\begin{align}
\Re\,\int_{-\infty}^t\int_{V} {\bf E}^{\text{T}}\frac{\partial {\bf P}}{\partial t^\prime}+{\bf H}^{\text{T}}\frac{\partial {\bf M}}{\partial t^\prime}\, d{\bf r} dt^\prime=\Im\, \frac{e^{-2i\omega t}}{-2i\omega}(F^\ast, h(\omega)F).
\end{align}
for all $t<T$. From this identity and (\ref{dynm_energy_den_real_fields}) it follows that
\begin{align}\label{dyn_energy_density_far_in_past_real_special_fields}
u_{\text{real}}(t) &= \frac{1}{2}u(t)+\frac{1}{16\pi}\Re\,e^{-2i\omega t}(F^\ast, F)\nonumber\\
&+\frac{1}{2}\Im\, \frac{e^{-2i\omega t}}{-2i\omega}(F^\ast, h(\omega)F)
\end{align}
for all $t<T$, where $u(t)$ is given in terms of $F$ by (\ref{dyn_energy_density_far_in_past_special_fields}). 

Now though, if $\omega_0\in (\omega_1,\omega_2)$, $\omega_0\not=0$, and $(\omega_1,\omega_2)\subseteq \mathbb{R}$ is a transparency window then it follows from (\ref{dyn_energy_density_far_in_past_special_fields}), (\ref{Im_part_herglotz_func_connect_to_deriv}), and (\ref{dyn_energy_density_far_in_past_real_special_fields}) that for $\omega=\omega_0+i\eta$, $\eta>0$ we have
\begin{align}\label{dynm_energy_density_limit_real_fields}
\lim_{\eta\downarrow 0}u_{\text{real}}(t)&=\frac{1}{16\pi}\left[(F,F)+4\pi(F,h^\prime(\omega_0)F)\right]\nonumber\\
&+\frac{1}{16\pi}\Re\,e^{-2i\omega_0 t}(F^\ast, F)\nonumber\\
&+\frac{1}{2}\Im\, \frac{e^{-2i\omega_0 t}}{-2i\omega_0}(F^\ast, h(\omega_0)F),
\end{align}
provided $t<T$. Letting $\left<f\right>$ denote the time-average of a periodic function $f$ with period $2\pi/\omega_0$ over a time interval of length $2\pi/\omega_0$ contained in $(-\infty,T)$, we find that
\begin{align}\label{dynm_energy_density_limit_real_EMWaves}
\left<\lim_{\eta\downarrow 0}u_{\text{real}}(t)\right>=\frac{1}{16\pi}\left[(F,F)+4\pi(F,h^\prime(\omega_0)F)\right] \nonumber \\
=\left(F,\frac{1}{16\pi}\frac{d}{d\omega}\omega\left[ I+4\pi\widehat{\chi}(\omega)\right]\big|_{\omega=\omega_0}F\right).
\end{align}

For homogeneous media, (\ref{dynm_energy_density_limit_real_EMWaves}) is just the familiar Brillouin formula (\ref{DefEnergyDensHomog}) for homogeneous media and, for periodic media, it is the generalization using the inner product $(\cdot,\cdot)$ defined in (\ref{Def:HilbertSpaceInnerProduct}).
In particular, if we had a time-harmonic Bloch wave in the form (\ref{dynm_energy_BlochWave_fields}), then the formula (\ref{dynm_energy_density_limit_real_EMWaves}) just becomes
\begin{align}\label{dynm_energy_density_limit_real_BlochWaves}
\left<\lim_{\eta\downarrow 0}u_{\text{real}}(t)\right>=\int_V U({\bf r})\,d{\bf r},
\end{align}
where $U$ is defined in (\ref{DefEnergyDensityGeneralCase}).

\section{Conclusion}\label{Sec:Conclusion}

In this paper we have generalized the notion of linear media and shown that periodic passive linear media implies the following: the polarization response obeys causality, i.e., the electromagnetic susceptibility $\chi(t)$ (up to an additive constant which we argue is zero on physical grounds) satisfies $\chi(t)=0$ for $t<0$ (in the distributional sense), existence of its Fourier transform $\widehat{\chi}(\omega)$ in the upper half-plane (i.e., $\Im\,\omega>0$), the function $h(\omega)=\omega\widehat{\chi}(\omega)$ is a bounded operator-valued Herglotz function (i.e., an analytic bounded-operator-valued function in the upper half-plane with a positive semidefinite imaginary part), and electromagnetic losses can be quantified in terms of a nonnegative-operator-valued measure $\Omega$ on the bounded Borel subsets of $\mathbb{R}$ defined in terms of a limit (as we approach the real axis from the upper half-plane) of an integral involving the imaginary part $\Im\,h(\omega)$. We have also shown that in a transparency window $(\omega_1,\omega_2)\subseteq \mathbb{R}$ (i.e., frequency window of negligible losses), which we defined in terms of that measure as $\Omega((\omega_1,\omega_2))=0$, the Herglotz function $h(\omega)$ has an analytic continuation into the lower-half plane through the transparency window such that $h(\omega)$ is a monotonically increasing (i.e., $h^\prime(\omega)\geq 0$) bounded-operator-valued function which is self-adjoint (i.e., $\Im\,h(\omega)=0$) in the window.  From this monotonicity property we showed that the energy density of any planewave or Bloch wave is always positive and the energy velocity ${\bf v}_e$ of such a wave is always bounded above by c (i.e., $||{\bf v}_e||\le c$), the speed-of-light in a vacuum. Furthermore, we showed that almost never is the upper bound $c$ achieved (i.e., $||{\bf v}_e||=c$), except in vacuum or in a few special cases, but if it is achieved then the dispersion relation $\widehat{\chi}(\omega)$ must be of a very specific form which is proportional to $1/\omega$, the latter of which is a very strong and unrealistic constraint upon the material dispersion since it typically decays faster then $1/\omega$ as $\omega\rightarrow \infty$. For lossy media, we generalized the result of Glasgow \textit{et al.}~\cite{Glas01} on the dynamical energy velocity ${\bf v}_e(t)$ being subluminal in passive media to include susceptibilities that are distributions as well as periodic and nonlocal media. Moreover, we have shown that dynamical energy density $U(t)$ and velocity ${\bf v}_e(t)$ reduce, in the limit to a transparency window, to the familiar Brillouin formula for energy density and velocity.

Although much has been said in this paper on bounding the energy velocity by $c$, there are still some major open questions and problems. First, how should one handle spatial nonlocality across unit cell boundaries? This is clearly going to involve imposing some constraints on the type of nonlocality that would be allowed. Second, how should one determine which of the many different types of well-known velocities for light \cite{Brill60}, \cite{Smit78}, \cite{Bloc77}, \cite[\S 2.9]{Milo05} such as the front velocity \cite{Brill60}, \cite[\S 5.18]{Strat07}, \cite[\S 2.1]{Milo05} (sometimes referred to as the signal velocity, see \cite{Smit78}) are also bounded by $c$ in passive linear media? Certain progress has been made on the front velocity, for instance, it is known that the total dynamical energy density in passive anisotropic local media has a front velocity bound by $c$ \cite{Glas01}. Moreover, since the work of Sommerfeld and Brillouin \cite{Sommerfeld1907, Sommerfeld1914, Brillouin1914, Brill60} on signal velocity (for a modern definition of a ``signal'' see, for instance, \cite[\S 2.9; \S 4.5]{Milo05}, \cite[\S 8]{Chiao97}), there has been significant mathematical development in the theory of Hyperbolic PDEs relating to the wave-front velocity, propagation of singularites, and finite speed of propagation (see, for instance, \cite{Rauch12}) which may be effective in the study of subliminal velocities of light in dissipative and dispersive linear media. Finally, how should one derive a similar type of velocity bounds, if possible, in nonlinear or time-varying media?

\begin{acknowledgments}
This work was supported in part by the Army Research Office through the Institute for Soldier Nanotechnologies under Contract No. W911NF-07-D0004. We would like to thank both Alex Figotin and Stephen Shipman for the many fruitful discussions relating to this paper.
\end{acknowledgments}

\appendix
\section{Energy Velocity and Group Velocity of Electromagnetic Bloch Waves in Periodic Media}\label{AppendixEnergyVelEqsGroupVel}

In this appendix we study the relationship between the energy velocity and group velocity of electromagnetic Bloch waves in periodic passive linear media in a transparency window, and show they are the same in local media. 

The equality of energy velocity and group velocity in media with negligible loss was previously shown for homogeneous dispersive media \cite{Abraham11}, \cite[Chap.~VI.31]{Havelock14}, \cite[\S IV.5]{Brill60}, \cite[\S 7.1]{Jack75}, \cite[\S 83]{Land84}, \cite{Yaghjian09}; for periodic non-dispersive media \cite[Appendix B]{Yeh79}, \cite[\S 8.5]{Jack75}, \cite[p.~41]{John08}; for periodic dispersive media \cite{Biot57, Wagner59I, Wagner59II, Bertoni66, McPhedran10}; and for spatial-temporal periodic media \cite{ChuTamir71}.

In this appendix, we combine and generalize these results to include some cases that were not handled previously, including periodic bianisotropic media and susceptibilities that are generalized functions. We also discuss the case of nonlocal media (within the unit cell), for which energy and group velocity are not in general equal. 
Our extension of previous results is straightforward, essentially using the variational approach of \cite{Wagner59I, Wagner59II}, \cite[Appendix B]{Yeh79}, \cite[p.~41]{John08} based on a generalization of the Hellmann--Feynman theorem of quantum mechanics \cite{Feynman39}, \cite{Hellmann37}, \cite[\S 11, Eq.~(11.16)]{Landau91Quantum}, \cite[Theorem 7.3.1]{Ismail05} which in essence is just a result of first-order perturbation theory for self-adjoint linear operators \cite{Epstein54, Musher66, Levine66, Pupyshev00}.

In the analysis that follows we assume the periodic media is passive in a periodic unit cell $V$ with a transparency window $(\omega_1,\omega_2)\subseteq \mathbb{R}$ and restrict ourselves to frequencies $\omega$ in this range. For a unified treatment, the unit cell $V$ is a measurable set which may be unbounded in directions that are not periodic (as in the case of a periodic waveguide), and may be lower-dimensional in cases of continuous translational invariance. For instance, if there are certain directions of continuous translational invariance (such as for homogeneous media), we can treat these as having ``period'' zero (reducing the dimensionality of $V$).

We consider time-harmonic electromagnetic Bloch waves
\begin{eqnarray}\label{DefEMBlochWaveAppendix}
{\bf E}={\bf E}_{\bf k}({\bf r})e^{i({\bf k}\cdot{\bf r}-\omega t)},\;\;
{\bf H}={\bf H}_{\bf k}({\bf r})e^{i({\bf k}\cdot{\bf r}-\omega t)},
\end{eqnarray}
where ${\bf E}_{\bf k}({\bf r})$, ${\bf H}_{\bf k}({\bf r})$ are periodic functions on the lattice, with real frequency $\omega\in(\omega_1,\omega_2)$ and real wavevector ${\bf k}\in\mathbb{R}^3$ (for $1d$ or $2d$ periodic media, the components of ${\bf k}$ are zero in non-periodic directions) such that $\begin{bmatrix} {\bf E}_{\bf k} & {\bf H}_{\bf k}\end{bmatrix}^{\text{T}}\in (L^2(V))^6$ (or $\mathbb{C}^6$ in the case of homogeneous media). For certain nonzero frequencies $\omega=\omega({\bf k})$, the Bloch waves are solutions of Maxwell's equations in the frequency domain
\begin{align}\label{MaxwellEqFreqDom}
\begin{bmatrix}
0 & \nabla\times\\
-\nabla\times & 0
\end{bmatrix}
\begin{bmatrix}
{\bf E}\\
{\bf H}
\end{bmatrix}
=-\frac{i\omega}{c}
\begin{bmatrix}
{\bf D}\\
{\bf B}
\end{bmatrix},\;\;\;
\begin{bmatrix}
{\bf D}\\
{\bf B}
\end{bmatrix}
=\begin{bmatrix}
{\bf E}\\
{\bf H}
\end{bmatrix}+4\pi \begin{bmatrix}
{\bf P}\\
{\bf M}
\end{bmatrix},
\end{align}
where $\bf P$ and $\bf M$ are given in terms of the susceptibility by
\begin{eqnarray}\label{SusceptibilityFreqDom}
\begin{bmatrix}
{\bf P}\\
{\bf M}
\end{bmatrix}=\widehat{\chi}(\omega)
\begin{bmatrix}
{\bf E}\\
{\bf H}
\end{bmatrix}
\end{eqnarray}
(the conditions $\nabla \cdot {\bf D}=0$, $\nabla \cdot {\bf B}=0$ are automatically satisfied since $\nabla \cdot\nabla\times=0$ should hold for the electromagnetic fields being considered.)

On the Hilbert space $\mathfrak{H}=(L^2(V))^6$ with inner product $(\cdot,\cdot)$ given by
\begin{eqnarray}\label{Def:HilbertSpaceInnerProduct}
(\psi,\varphi)=\int_V\psi({\bf r})^\dagger\varphi({\bf r})d{\bf r},\;\;\psi,\varphi\in \mathfrak{H}
\end{eqnarray}
(or, in the case of homogeneous media, $\mathfrak{H}=\mathbb{C}^6$ with inner product $(\cdot,\cdot)$ given by $(\psi,\varphi)=\psi^\dagger\varphi$), the operator $h(\omega)=\omega\widehat{\chi}(\omega)$, because of the hypotheses of passivity and transparency window, has the properties that it is a bounded-operator-valued function which is analytic in a region containing $(\omega_1,\omega_2)$ and self-adjoint for any $\omega\in (\omega_1,\omega_2)$ satisfying $h^{\prime}(\omega)\geq 0$ for every $\omega\in (\omega_1,\omega_2)$. We introduce the operator $\mathcal{T}({\bf k})$ defined on $\mathfrak{H}$ by multiplication
\begin{eqnarray}\label{DefMultiplicationOperatorExpAndWavenumber}
(\mathcal{T}({\bf k})\psi)({\bf r})=e^{i{\bf k}\cdot{\bf r}}\psi({\bf r}),\;\;\psi\in \mathfrak{H}.
\end{eqnarray}

For this type of geometry and these types of solutions, the curl operator  satisfies
\begin{align}
\nabla\times=\mathcal{T}({\bf k})\mathcal{A}({\bf k})\mathcal{T}(-{\bf k}),\text{ where }\mathcal{A}({\bf k})=\nabla\times+i{\bf k}\times
\end{align}
Thus, the equations (\ref{MaxwellEqFreqDom}), (\ref{SusceptibilityFreqDom}) become
\begin{widetext}
\begin{gather}\label{MaxwellEqFreqDomBlochPeriodicPart}
\mathcal{M}(\omega,{\bf k})
\begin{bmatrix}
{\bf E}_{\bf k}\\
{\bf H}_{\bf k}
\end{bmatrix}={\bf 0},\;\;\text{where }\mathcal{M}(\omega,{\bf k})=\mathcal{M}_0+\mathcal{M}_1(\omega,{\bf k}),\\
\mathcal{M}_0=\begin{bmatrix}
0 & -\frac{ic}{16\pi}\nabla\times\\
\frac{ic}{16\pi}\nabla\times & 0
\end{bmatrix},\;\mathcal{M}_1(\omega,{\bf k})=\begin{bmatrix}
0 & \frac{c}{16\pi}{\bf k}\times\\
-\frac{c}{16\pi}{\bf k}\times & 0
\end{bmatrix}+\frac{\omega}{16\pi}I+\frac{1}{4\pi}\mathcal{T}(-{\bf k})h(\omega)\mathcal{T}({\bf k}).
\end{gather}
\end{widetext}

Suppose now that at some real point $(\omega_0,{\bf k}_0)$ in $(\omega_1,\omega_2)\times \mathbb{R}^3$ and for variations ${\bf k}={\bf k}_0+\delta{\bf k}$, $||\delta{\bf k}||\ll 1$, along the periodic directions that there exists a continuous function $\omega=\omega({\bf k})$ and family of Bloch wave solutions to (\ref{MaxwellEqFreqDomBlochPeriodicPart}) satisfying $\omega=\omega_0+o(1)$, $\begin{bmatrix} {\bf E}_{\bf k} & {\bf H}_{\bf k}\end{bmatrix}^{\text{T}}=\begin{bmatrix} {\bf E}_{{\bf k}_0} & {\bf H}_{{\bf k}_0}\end{bmatrix}^{\text{T}}+o(1)$ (in the norm topology of $\mathfrak{H}$) as $||\delta{\bf k}||\rightarrow 0$. We will now prove that the total differential $d\omega$ exists and derive a formula for the gradient
\begin{align}
\nabla_{\bf k}\omega({\bf k}_0)=\displaystyle\sum_{j}\frac{\partial \omega}{\partial k_j}({\bf k}_0)\,{\bf e}_j,
\end{align}
where ${\bf e}_i$, $i=1,2,3$ denote the standard basis vectors in $\mathbb{R}^3$ and the sum is over the $j$ in the periodic directions. For typical boundary conditions, the usual integration by parts formula for these types of periodic solutions to (\ref{MaxwellEqFreqDomBlochPeriodicPart}) hold, i.e.,
\begin{align}
\left(\mathcal{M}_0\begin{bmatrix}
{\bf E}_{\bf k}\\
{\bf H}_{\bf k}
\end{bmatrix},\begin{bmatrix}
{\bf E}_{{\bf k}_0}\\
{\bf H}_{{\bf k}_0}
\end{bmatrix}\right)-\left(\begin{bmatrix}
{\bf E}_{\bf k}\\
{\bf H}_{\bf k}
\end{bmatrix},\mathcal{M}_0\begin{bmatrix}
{\bf E}_{{\bf k}_0}\\
{\bf H}_{{\bf k}_0}
\end{bmatrix}\right)=0,
\end{align}
and so it follows by our assumptions that
\begin{align}
&\left(\mathcal{M}(\omega,{\bf k})
\begin{bmatrix}
{\bf E}_{\bf k}\\
{\bf H}_{\bf k}
\end{bmatrix},\begin{bmatrix}
{\bf E}_{{\bf k}_0}\\
{\bf H}_{{\bf k}_0}
\end{bmatrix}\right)-\left(
\begin{bmatrix}
{\bf E}_{\bf k}\\
{\bf H}_{\bf k}
\end{bmatrix},\mathcal{M}(\omega_0,{\bf k}_0)\begin{bmatrix}
{\bf E}_{{\bf k}_0}\\
{\bf H}_{{\bf k}_0}
\end{bmatrix}\right)\nonumber\\
&=\left(
\begin{bmatrix}
{\bf E}_{\bf k}\\
{\bf H}_{\bf k}
\end{bmatrix},\delta\mathcal{M}_1\begin{bmatrix}
{\bf E}_{{\bf k}_0}\\
{\bf H}_{{\bf k}_0}
\end{bmatrix}\right)=0,
\end{align}
where
\begin{gather}
\delta\mathcal{M}_1=\mathcal{M}_1(\omega,{\bf k})-\mathcal{M}_1(\omega_0,{\bf k}_0)\nonumber\\
=\delta\omega\frac{\displaystyle\partial\mathcal{M}_1}{\displaystyle\partial \omega}(\omega_0,{\bf k}_0)+\sum_{j}\delta k_j\frac{\displaystyle\partial\mathcal{M}_1}{\displaystyle\partial k_j}(\omega_0,{\bf k}_0)+o(\delta\omega)+o(||\delta{\bf k}||)
\end{gather}
(in the operator norm topology of $[\mathfrak{H};\mathfrak{H}]$ -- the space of continuous linear operators on $\mathfrak{H}$) as $||\delta{\bf k}||\rightarrow 0$. From this, our variation assumptions, and the Schwarz inequality \cite{Sim81} for the inner product $(\cdot,\cdot)$ it follows that
\begin{align}
&0=\left(
\begin{bmatrix}
{\bf E}_{\bf k}\\
{\bf H}_{\bf k}
\end{bmatrix},\delta\mathcal{M}_1\begin{bmatrix}
{\bf E}_{{\bf k}_0}\\
{\bf H}_{{\bf k}_0}
\end{bmatrix}\right)\nonumber\\
&=\left(
\begin{bmatrix}
{\bf E}_{{\bf k}_0}\\
{\bf H}_{{\bf k}_0}
\end{bmatrix},\delta\mathcal{M}_1\begin{bmatrix}
{\bf E}_{{\bf k}_0}\\
{\bf H}_{{\bf k}_0}
\end{bmatrix}\right)+\left(
\delta\begin{bmatrix}
{\bf E}_{\bf k}\\
{\bf H}_{\bf k}
\end{bmatrix},\delta\mathcal{M}_1\begin{bmatrix}
{\bf E}_{{\bf k}_0}\\
{\bf H}_{{\bf k}_0}
\end{bmatrix}\right)\nonumber\\
&=\delta\omega\left(
\begin{bmatrix}
{\bf E}_{{\bf k}_0}\\
{\bf H}_{{\bf k}_0}
\end{bmatrix},\frac{\displaystyle\partial\mathcal{M}_1}{\displaystyle\partial \omega}(\omega_0,{\bf k}_0)
\begin{bmatrix}
{\bf E}_{{\bf k}_0}\\
{\bf H}_{{\bf k}_0}
\end{bmatrix}\right)\nonumber\\
&+\sum_{j}\delta k_j\left(
\begin{bmatrix}
{\bf E}_{{\bf k}_0}\\
{\bf H}_{{\bf k}_0}
\end{bmatrix},\frac{\displaystyle\partial\mathcal{M}_1}{\displaystyle\partial k_j}(\omega_0,{\bf k}_0)
\begin{bmatrix}
{\bf E}_{{\bf k}_0}\\
{\bf H}_{{\bf k}_0}
\end{bmatrix}\right)+o(\delta\omega)+o(||\delta{\bf k}||)
\end{align}
as $||\delta{\bf k}||\rightarrow 0$. It follows from this and the fact
\begin{align}\label{DefEnergyDensityGeneralCaseSupplement}
\frac{\partial\mathcal{M}_1}{\partial \omega}(\omega_0,{\bf k}_0)=\frac{1}{16\pi}I+\frac{1}{4\pi}\mathcal{T}(-{\bf k}_0)h^\prime(\omega_0)\mathcal{T}({\bf k}_0)\geq \frac{1}{16\pi}I, 
\end{align} 
that the total differential $d\omega$ exists and is given by the formula
\begin{align}
d\omega=\sum_{j}\frac{-\left(
\begin{bmatrix}
{\bf E}_{{\bf k}_0}\\
{\bf H}_{{\bf k}_0}
\end{bmatrix},\frac{\displaystyle\partial\mathcal{M}_1}{\displaystyle\partial k_j}(\omega_0,{\bf k}_0)
\begin{bmatrix}
{\bf E}_{{\bf k}_0}\\
{\bf H}_{{\bf k}_0}
\end{bmatrix}\right)}{\left(
\begin{bmatrix}
{\bf E}_{{\bf k}_0}\\
{\bf H}_{{\bf k}_0}
\end{bmatrix},\frac{\displaystyle\partial\mathcal{M}_1}{\displaystyle\partial \omega}(\omega_0,{\bf k}_0)
\begin{bmatrix}
{\bf E}_{{\bf k}_0}\\
{\bf H}_{{\bf k}_0}
\end{bmatrix}\right)}dk_j.
\end{align}
From the vector identity
\begin{align}
\Re \left({\bf E}\times{\bf H}^\ast\right)=-\frac{1}{2}\displaystyle\sum_{j=1}^3\begin{bmatrix}
{\bf E}\\
{\bf H}
\end{bmatrix}^\dagger
\begin{bmatrix}
0 & {\bf e}_j\times\\
-{\bf e}_j\times & 0
\end{bmatrix}
\begin{bmatrix}
{\bf E}\\
{\bf H}
\end{bmatrix}\,{\bf e}_j,
\end{align}
it follows that the numerator is given by the formula
\begin{align}\label{GrpVelNumerForm}
&-\left(\begin{bmatrix}
{\bf E}_{{\bf k}_0}\\
{\bf H}_{{\bf k}_0}
\end{bmatrix},\frac{\displaystyle\partial\mathcal{M}_1}{\partial k_j}(\omega_0,{\bf k}_0)\begin{bmatrix}
{\bf E}_{{\bf k}_0}\\
{\bf H}_{{\bf k}_0}
\end{bmatrix}\right)\nonumber\\
&=\int_V \Re\, {\bf S}({\bf r})\cdot {\bf e}_j d{\bf 
r}-4\pi\left(\begin{bmatrix}
{\bf E}_{{\bf k}_0}\\
{\bf H}_{{\bf k}_0}
\end{bmatrix},\frac{\partial h}{\partial k_j}(\omega_0,{\bf k}_0)\begin{bmatrix}
{\bf E}_{{\bf k}_0}\\
{\bf H}_{{\bf k}_0}
\end{bmatrix}\right),
\end{align}
where ${\bf S}=\frac{c}{8\pi}{\bf E}_{{\bf k}_0}\times{{\bf H}^\ast_{{\bf k}_0}}$ is the spatially dependent complex Poynting vector and
\begin{eqnarray}
h(\omega,{\bf k})=\mathcal{T}(-{\bf k})h(\omega)\mathcal{T}({\bf k}).
\end{eqnarray}
By the formula for the total differential $d\omega$ it follows immediately that
\begin{align}\label{GrpVelViaHellmFeyn}
\frac{\partial \omega}{\partial k_j}({\bf k}_0)=\frac{-\left(\begin{bmatrix}
{\bf E}_{{\bf k}_0}\\
{\bf H}_{{\bf k}_0}
\end{bmatrix},\frac{\displaystyle\partial\mathcal{M}_1}{\displaystyle\partial k_j}(\omega_0,{\bf k}_0)\begin{bmatrix}
{\bf E}_{{\bf k}_0}\\
{\bf H}_{{\bf k}_0}
\end{bmatrix}\right)}{\left(\begin{bmatrix}
{\bf E}_{{\bf k}_0}\\
{\bf H}_{{\bf k}_0}
\end{bmatrix},\frac{\displaystyle\partial\mathcal{M}_1}{\displaystyle\partial \omega}(\omega_0,{\bf k}_0)\begin{bmatrix}
{\bf E}_{{\bf k}_0}\\
{\bf H}_{{\bf k}_0}
\end{bmatrix}\right)}.
\end{align}

For local media, $\mathcal{T}({\bf k})$ and $h(\omega)$ commute so that the second term on the RHS of (\ref{GrpVelNumerForm}) is zero (for generic non-local media this will not be true) and hence it follows from the derived equalities that the group velocity is given by the formulas
\begin{align}\label{IdentityGrpVelEqsEnergyVelGenerally}
\frac{\partial \omega}{\partial k_j}({\bf k}_0)=\frac{\int_V \Re\, {\bf S}({\bf r})\cdot {\bf e_j} d{\bf r}}{\int_V U({\bf r})d{\bf r}},
\end{align}
where
\begin{gather}
{\bf S}=\frac{c}{8\pi}{\bf E}_{{\bf k}_0}\times{{\bf H}^\ast_{{\bf k}_0}}\\
U=\begin{bmatrix}\label{DefEnergyDensityGeneralCase}
{\bf E}_{{\bf k}_0}\\
{\bf H}_{{\bf k}_0}
\end{bmatrix}^\dagger\frac{\partial\mathcal{M}_1}{\partial \omega}(\omega_0,{\bf k}_0)\begin{bmatrix}
{\bf E}_{{\bf k}_0}\\
{\bf H}_{{\bf k}_0}
\end{bmatrix},
\end{gather}
where $\frac{\partial\mathcal{M}_1}{\partial \omega}(\omega_0,{\bf k}_0)$ is given by the formula (\ref{DefEnergyDensityGeneralCaseSupplement}) and (\ref{DefMultiplicationOperatorExpAndWavenumber}).

\section{Proofs}\label{AppendixSecProofs}
In this appendix we prove the rest of statements in this paper from Sec.~\ref{Sec:HomogPassivity}, \ref{Sec:HomogTranspWnd}, \ref{Sec:PeriodicPassivity}, \ref{Sec:PeriodicTranspWnd}, and \ref{Sec:Achieving_c} that were left to be proved here.

This section is organized as follows. First, in Sec.~\ref{AppdxSecNotationsConventionsDefinitions} we give our notations, conventions, and definitions that will be needed in our proofs. Next, in Sec.~\ref{AppdxOnAntiderivativesOfCertainDistributions} we state and prove two results on antiderivatives of certain classes of distributions which will be needed in the next section. Finally, Sec.~\ref{AppdxSecProofsOfMainResults} contains the main body of this appendix, namely, the proofs of the statements in this paper from the sections mentioned above.

\subsection{Notations, Conventions, and Definitions}\label{AppdxSecNotationsConventionsDefinitions}
Unless otherwise indicated, we will adhere to the notation, conventions, and definitions in \cite{Zema72}. The following modifications and additional definitions and notations should be noted. 

First, any Hilbert space will be denoted by $\mathfrak{H}$ with the convention that the inner product $(\cdot,\cdot)$ is antilinear in first vector coordinate rather then the second. As in \cite[p.~3]{Zema72}, $[\mathfrak{H};\mathfrak{H}]$ denotes the space of all continuous linear operators on $\mathfrak{H}$. 

Second, we will use the notation $\mathbb{R}$ to denote the set of real numbers as opposed to $R$. Thus, as in \cite[p.~50 \& p.~52]{Zema72}, $\mathscr{D}(\mathfrak{H})$ denotes the space of all infinitely differentiable functions $\phi:\mathbb{R}\rightarrow \mathfrak{H}$ with compact support and $[\mathscr{D}(\mathfrak{H});\mathfrak{H}]$ denotes the $[\mathfrak{H};\mathfrak{H}]$-valued distributions (or generalized functions), that is, the set of all continuous linear functions $f:\mathscr{D}(\mathfrak{H})\rightarrow \mathfrak{H}$.

Third, as opposed to the definition in \cite[p.~66 \& \S 8.4]{Zema72}, the Fourier transform $\mathscr{F}$ of any $\phi\in \mathscr{S}(\mathfrak{H})$ (the $\mathfrak{H}$-valued testing functions of rapid descent) will be denoted by $\widehat \phi=\mathscr{F}\phi$ and defined by
\begin{align}
\widehat f (\omega)=(\mathscr{F}\phi)(\omega)=\int_{-\infty}^{\infty}\phi(t)e^{i\omega t}dt,\;\;\;\omega\in\mathbb{R}.\label{AppdxDefFourierTransformTestingFuncRapidDescent}
\end{align}

Fourth, using the definition in \cite[\S 6.2]{Zema72} of the Laplace transform on distributions belonging to the generalized function space $[\mathscr{L}(0,\infty;\mathfrak{H});\mathfrak{H}]$, the Fourier transform on such distributions can be defined. In particular, for a Laplace-transformable distribution $y\in [\mathscr{D}(\mathfrak{H});\mathfrak{H}]$ with strip of definition for the Laplace transform of $y$ containing $C_+=\{\zeta\in\mathbb{C}:\Re\,\zeta>0\}$ then this means $y\in[\mathscr{L}(0,\infty;\mathfrak{H});\mathfrak{H}]$ and it has a Laplace transform $\mathfrak{L}y:C_+\rightarrow [\mathfrak{H};\mathfrak{H}]$ as defined in \cite[\S 6.2, Eq.~(2)]{Zema72}. From the Laplace transform, the Fourier transform of $y$ denoted by $\widehat y=\mathscr{F}y$ is defined as the function
\begin{align}
\widehat y(\omega)=(\mathscr{F}y)(\omega)=(\mathfrak{L}y)(-i\omega),\;\;\;\omega\in\mathbb{C}_+,
\end{align}
where $\mathbb{C}_+=\{\omega\in\mathbb{C}:\Im\,\omega>0\}$.

Finally, a function $h:\mathbb{C}_+\rightarrow [\mathfrak{H};\mathfrak{H}]$ is said to be a Herglotz function if it is analytic on $\mathbb{C}_+$ such that $\Im\,h(\omega)\geq 0$ for every $\omega\in \mathbb{C}_+$, where $\Im$ denotes the imaginary part of an operator from $[\mathfrak{H};\mathfrak{H}]$ (i.e., for any $A\in [\mathfrak{H};\mathfrak{H}]$, $\Im\,A=\frac{1}{2i}(A-A^\dagger)$, where $A^\dagger$ denotes the adjoint of $A$).

\subsection{On Antiderivatives of Certain Distributions}\label{AppdxOnAntiderivativesOfCertainDistributions}
In this section we state and prove certain results on antiderivatives of distributions that will be needed in Sec.~\ref{AppdxSecProofsOfMainResults}. Although the results here seem to be known, we haven't been able to find references to them in the literature and hence for completeness we prove them here. The two key results below are (i) and (ii). The result (i) says that for a given $[\mathfrak{H};\mathfrak{H}]$-valued distribution $f$ with support in $[0,\infty)$, i.e., $f(t)=0$ for $t<0$ in the distributional sense \cite[p.~55]{Zema72}, there exists a unique antiderivative of $f$ with the same property. Result (ii) is almost identical to (i) except the hypothesis and conclusion that $f$ and its antiderivative are in $[\mathscr{D}(\mathfrak{H});\mathfrak{H}]$ is instead replaced by the subspace $[\mathscr{L}(0,\infty;\mathfrak{H});\mathfrak{H}]$.

To begin, recall that the generalized derivative $D:[\mathscr{D}(\mathfrak{H});\mathfrak{H}]\rightarrow [\mathscr{D}(\mathfrak{H});\mathfrak{H}]$ on distributions is defined as in \cite[p.~54]{Zema72} by
\begin{align}
\left<Df,\phi\right>=-\left<f,D\phi\right>,\;\;\;\phi\in \mathscr{D}(\mathfrak{H}).\label{AppdxDefGeneralizedDeriv}
\end{align}
Now an antiderivative (or primitive) of a distribution $f\in[\mathscr{D}(\mathfrak{H});\mathfrak{H}]$ is any $g\in[\mathscr{D}(\mathfrak{H});\mathfrak{H}]$ such that $Dg=f$. Following virtually the same argument from \cite[\S 5.6]{Zema72}, one can easily show that every distribution has an antiderivative and any two antiderivatives differ by a constant distribution (i.e., if $Dg_1=Dg_2$ for some $g_1,g_2\in[\mathscr{D}(\mathfrak{H});\mathfrak{H}]$ then there exists an $A\in[\mathfrak{H};\mathfrak{H}]$ such that $g_1-g_2=A$, where $\left<A,\phi\right>=\int_{-\infty}^{\infty}A\phi(t)dt$ for all $\phi\in \mathscr{D}(\mathfrak{H})$).

We will need the following two facts in this paper on antiderivatives of certain distributions which we will prove below: (i) if $f\in[\mathscr{D}(\mathfrak{H});\mathfrak{H}]$ and $\operatorname{supp} f\subseteq [0,\infty)$ then there exists a unique antiderivative $f^{(-1)}$ of $f$ such that $\operatorname{supp} f^{(-1)}\subseteq [0,\infty)$; (ii) if $f$ is a distribution in $[\mathscr{L}(0,\infty;\mathfrak{H});\mathfrak{H}]$ and $\operatorname{supp} f\subseteq [0,\infty)$ then there exists a unique antiderivative $f^{(-1)}$ of $f$ such that $f^{(-1)}\in[\mathscr{L}(0,\infty;\mathfrak{H});\mathfrak{H}]$ and $\operatorname{supp} f^{(-1)}\subseteq [0,\infty)$.

We prove these two statements now. We begin with the proof of statement (i) following the argument in \cite[\S 5.6]{Zema72} on the existence of an antiderivative. Suppose $f\in[\mathscr{D}(\mathfrak{H});\mathfrak{H}]$ and $\operatorname{supp} f\subseteq [0,\infty)$. Let $H$ denote the subspace of $\mathscr{D}(\mathfrak{H})$ whose elements $\varphi$ have the form $\varphi=D\psi$, where $\psi\in \mathscr{D}(\mathfrak{H})$. Let $\phi_0\in \mathscr{D}(\mathbb{C})$ be such that $\int_{-\infty}^{\infty}\phi_0(t)dt=1$ and $\operatorname{supp}\phi_0\subseteq (-\infty,0)$. Then every $\phi\in \mathscr{D}(\mathfrak{H})$ has the unique decomposition
\begin{align}
\phi=\phi_0w+\varphi\label{AppdxForAntiderivTestFuncDecomp}
\end{align}
where $w=\int_{-\infty}^{\infty}\phi(t)dt$ and $\varphi\in H$. In particular, $\varphi=D\psi$, where $\psi(t)=\int_{-\infty}^t\varphi(t)dt$ and $\psi\in \mathscr{D}(\mathfrak{H})$. Similar to \cite[\S 5.6, Eq.~(2)]{Zema72}, one defines a distribution $f^{(-1)}\in [\mathscr{D}(\mathfrak{H});\mathfrak{H}]$ in terms of the decomposition (\ref{AppdxForAntiderivTestFuncDecomp}) of $\phi\in \mathscr{D}(\mathfrak{H})$ by
\begin{align}
\left<f^{(-1)},\phi\right>=-\left<f,\psi\right>.\label{AppdxAntiderivDistrSuppOnRight}
\end{align}
From this definition and (\ref{AppdxDefGeneralizedDeriv}), it follows immediately that $f^{(-1)}$ is an antiderivative of $f$ and if $\phi\in \mathscr{D}(\mathfrak{H})$ and $\operatorname{supp} \phi\subseteq (-\infty,0)$ then $\operatorname{supp} \psi\subseteq (-\infty,0)$ so that since $\operatorname{supp} f\subseteq [0,\infty)$ this implies $0=-\left<f,\psi\right>=\left<f^{(-1)},\phi\right>$. This proves that $\operatorname{supp} f^{(-1)}\subseteq [0,\infty)$. The proof of uniqueness now just follows immediately from the facts two antiderivatives of a distribution differ by a constant distribution and the only constant distribution $A$ with $\operatorname{supp} A\subseteq [0,\infty)$ is $A=0$. This completes the proof of statement (i).

We now prove statement (ii). Suppose $f$ is a distribution in $[\mathscr{L}(0,\infty;\mathfrak{H});\mathfrak{H}]$ and $\operatorname{supp} f\subseteq [0,\infty)$. Then it follows from \cite[Thm 6.4-2; Thm 6.5-1]{Zema72} and \cite[Appendix E, E2]{Zema72} that there exists a unique Laplace-transformable distribution $f^{(-1)}$ in $[\mathscr{L}(0,\infty;\mathfrak{H});\mathfrak{H}]$ with $\operatorname{supp} f^{(-1)}\subseteq [0,\infty)$ such that its Laplace-transform $\mathfrak{L}f^{(-1)}:C_+\rightarrow [\mathfrak{H};\mathfrak{H}]$ is given by
\begin{align}
(\mathfrak{L}f^{(-1)})(\zeta)=\frac{1}{\zeta}(\mathfrak{L}f)(\zeta),\;\;\;\zeta\in C_+.
\end{align}
It follows immediately from this and \cite[Prob.~6.3-2]{Zema72} that $(\mathfrak{L}Df^{(-1)})(\zeta)=(\mathfrak{L}f)(\zeta)$ for every $\zeta\in C_+$ which, by the uniqueness of the Laplace transform \cite[Thm 6.4-2]{Zema72}, implies $f^{(-1)}$ is an antiderivative of $f$. This proves the existence portion of statement (ii). The uniqueness portion follows immediately from statement (i). This completes the proof of statement (ii).

\subsection{Proofs of Main Results}\label{AppdxSecProofsOfMainResults}
This section contains the main body of this appendix. Here we prove the statements from Sec.~\ref{Sec:HomogPassivity}, \ref{Sec:HomogTranspWnd}, \ref{Sec:PeriodicPassivity}, \ref{Sec:PeriodicTranspWnd}, and \ref{Sec:Achieving_c} that were left to be proved.  Our proofs are essentially based on a translation of the electromagnetic assumptions of passive linear media and transparency window into the distributional approach of Zemanian on the theory of passive linear systems \cite[Chap.~8]{Zema72} and connecting all of this to the theory of Herglotz functions (see, for instance, \cite{Aron56}, \cite{Gesz00}, \cite{Gesz01}, \cite[Appendix A]{Gesz13}, and references within). 

This section is organized as follows. We begin in Sec.~\ref{AppdxSecSusceptibilityAndPassiveLinearMedia} by recasting the polarization response (\ref{DefSusceptibConvol}) and passivity, either (\ref{DefPassiveHomog}) or (\ref{DefPassiveInhomog}), in terms of the theory of distributions and passive linear systems from \cite[Chap.~8]{Zema72} using the susceptibility $\chi$, a generalized function. Then in Sec.~\ref{AppdxSecProofsofSecHomogPassivityAndSecPeriodicPassivity} we prove the statements from Sec.~\ref{Sec:HomogPassivity} and \ref{Sec:PeriodicPassivity}, in particular, that linearity (\ref{DefSusceptibConvol}) and passivity (\ref{DefPassiveHomog}), (\ref{DefPassiveInhomog}) implies the three main consequences of causality, analyticity, and positivity. Our proof though essentially follows from \cite[Thm 8.12-1]{Zema72}, \cite[Lemma 7.2-1]{Zema72}, the existence of antiderivatives of distributions as discussed in Sec.~\ref{AppdxOnAntiderivativesOfCertainDistributions}, the relationship between the Laplace and Fourier transforms as discussed in Sec.~\ref{AppdxSecNotationsConventionsDefinitions}, and the relationship between positive* mappings \cite[Def.~8.11-1]{Zema72} and Herglotz functions (defined in Sec.~\ref{AppdxSecNotationsConventionsDefinitions}). Finally, in Sec.~\ref{AppdxSecProofsofSecHomogTranspWndSecPeriodicTranspWndAndSecAchievingc} we prove the rest of statements that were left to be proved here from \ref{Sec:HomogTranspWnd}, \ref{Sec:PeriodicTranspWnd}, and \ref{Sec:Achieving_c}. In particular, we give a precise and rigorous definition of a transparency window using the Herglotz function $h(\omega)=\omega\widehat \chi(\omega)$ from (\ref{DefHerglotzFuncHomog}) and prove the main consequences of a transparency window $(\omega_1,\omega_2)$, namely, analytic continuation of $h(\omega)$ across $(\omega_1,\omega_2)$ to a bounded-operator-valued function satisfying the self-adjoint and monotonicity conditions (\ref{HerglotzHomogMonot}). We also prove in Sec.~\ref{AppdxSecProofsofSecHomogTranspWndSecPeriodicTranspWndAndSecAchievingc} the fact mentioned in Sec.~\ref{Sec:Achieving_c} on achieving the upper bound $||{\bf v}_e||=c$ that the condition (\ref{HerglotzPlanewaveNullspace}) being true for $h$ at a single frequency in a transparency window implies that (\ref{HerglotzPlanewaveNullspace}) and (\ref{HerglotzPlanewaveNullspace1}) are true for all frequencies. Our proofs of all these results are essentially just based on the connection between the well-known integral representation of a Herglotz function from \cite[Theorem 5.4]{Gesz00}, \cite[Eqs.~(1.1)--(1.3)]{Gesz01}, and \cite[Theorem A.4]{Gesz13} to the Schwindt's representation of a positive* mapping \cite[Theorem 8.11-2]{Zema72} and the characterization of passive convolution operators in terms of positive* mappings \cite[Theorem 8.12-1]{Zema72}. 

\subsubsection{Susceptibility and Passive Linear Media}\label{AppdxSecSusceptibilityAndPassiveLinearMedia}
In this section we show how the electromagnetic assumptions of passive linear media are related to the theory of distributions and the theory of passive linear systems \cite[Chap.~8]{Zema72} through the susceptibility $\chi$, a generalized function, and its time derivative $\frac{d\chi}{dt}$ (in the distributional sense as defined in \cite[p.~54]{Zema72} or as in Sec.~\ref{AppdxOnAntiderivativesOfCertainDistributions}, (\ref{AppdxDefGeneralizedDeriv}) of this paper).

To begin with, the Hilbert spaces we are dealing with are either $\mathfrak{H}=\mathbb{C}^6$ with inner product $(\psi,\varphi)=\psi^\dagger \varphi$ or $\mathfrak{H}=(L^2(V))^6$ for some measurable set $V\subseteq\mathbb{R}^d$, $d=1,2,$ or $3$, with inner product $(\cdot,\cdot)$ given by
\begin{eqnarray}\label{AppdxDefHilbertSpaceInnerProduct}
(\psi,\varphi)=\int_V\psi({\bf r})^\dagger\varphi({\bf r})d{\bf r},\;\;\psi,\varphi\in \mathfrak{H},
\end{eqnarray}
Also, for $\mathfrak{H}=\mathbb{C}^6$, we may replace in the following sections any limit in the strong operator topology ($s\mbox{-}\lim$) with the limit in the operator norm topology ($\lim$) since in any finite-dimensional Hilbert space the operator norm topology and strong operator topology (and weak operator topology) are equivalent \cite[Chap.~IV]{Halmos74}, \cite[\S VI.1]{Sim81}. 

Now the susceptibility $\chi$ is a generalized function (distribution), i.e., $\chi\in [\mathscr{D}(\mathfrak{H});\mathfrak{H}]$, and polarization response (\ref{DefSusceptibConvol}) is given in terms of the convolution operator $\chi \ast:\mathscr{D}(\mathfrak{H})\rightarrow \mathscr{E}(\mathfrak{H})$ (as defined in \cite[Lemma 5.5-1, Eq.~(1)]{Zema72}) by $(\chi \ast v)(t)=\left<\chi,v(t-\cdot)\right>$. Thus, since $D(\chi\ast v)=(D\chi)\ast v$ for every $v\in \mathscr{D}(\mathfrak{H})$ (where $D=\frac{d}{dt}$ denotes the generalized derivative) which follows from \cite[Lemma 5.5-1, Eq.~(2)]{Zema72} and the definition (\ref{AppdxDefGeneralizedDeriv}), it follows that the main assumption in our paper of passivity, either (\ref{DefPassiveHomog}) or (\ref{DefPassiveInhomog}), is exactly the condition
\begin{align}
0\leq\Re\,\int_{-\infty}^t(v(\tau), (D\chi)\ast v\,(\tau))d\tau\label{AppdxProofOfMainResultsPassivityCondition}
\end{align}
for every $v\in \mathscr{D}(\mathfrak{H})$ and every $t\in\mathbb{R}$. Hence, by definition \cite[Def.~8.2-2]{Zema72}, the condition (\ref{AppdxProofOfMainResultsPassivityCondition}) means that $D\chi\ast$ is a passive convolution operator on $\mathscr{D}(\mathfrak{H})$. 

\subsubsection{Proofs of Results in Sec.~\ref{Sec:HomogPassivity}, \ref{Sec:PeriodicPassivity}}\label{AppdxSecProofsofSecHomogPassivityAndSecPeriodicPassivity}
In this section we prove the statements from Sec.~\ref{Sec:HomogPassivity} and \ref{Sec:PeriodicPassivity}, namely, that linearity (\ref{DefSusceptibConvol}) and passivity (\ref{DefPassiveHomog}), (\ref{DefPassiveInhomog}) implies causality, analyticity, and positivity as described in Sec.~\ref{Sec:HomogPassivity}, \ref{Sec:PeriodicPassivity}. As we will show, these results just follow from the facts that $D\chi\ast$ is a passive convolution operator on $\mathscr{D}(\mathfrak{H})$, $\chi$ is an antiderivative of $D\chi$, and our existence and uniqueness results in Sec.~\ref{AppdxOnAntiderivativesOfCertainDistributions} on antiderivatives of distributions with support in $[0,\infty)$. Then we conclude this section with two results that completely characterize $D\chi\ast$ as a passive convolution operator in terms of the Herglotz function $h(\omega)=\omega\widehat \chi(\omega)$ from (\ref{DefHerglotzFuncHomog}).

Now to begin, we have already shown in Sec.~\ref{AppdxSecSusceptibilityAndPassiveLinearMedia} that linearity (\ref{DefSusceptibConvol}) and passivity (\ref{DefPassiveHomog}), (\ref{DefPassiveInhomog}) imply $D\chi\ast$ is a passive convolution operator on $\mathscr{D}(\mathfrak{H})$. This implies by \cite[Thm 8.2-1]{Zema72} that $D\chi\ast$ is a causal convolution operator on $\mathscr{D}(\mathfrak{H})$ \cite[Def.~4.6-1]{Zema72} and hence \cite[Thm 5.11-1]{Zema72} implies that $(D\chi)(t)=0$ for all $t<0$, that is, $\operatorname{supp}(D\chi)\subseteq [0,\infty)$. From this property it follows from our results in Sec.~\ref{AppdxOnAntiderivativesOfCertainDistributions} that there exists a unique antiderivative of $D\chi$, say $X$, such that $\operatorname{supp}X\subseteq [0,\infty)$. It then follows from \cite[Thm 5.11-1]{Zema72} that $X\ast$ is a causal convolution operator on $\mathscr{D}(\mathfrak{H})$. As we discussed in Sec.~\ref{AppdxOnAntiderivativesOfCertainDistributions}, since $\chi$ is also an antiderivative of $D\chi$, this implies $\chi-X$ must be a constant distribution. Our proofs below show that, in fact, we do not require this constant to be zero, but in order to simplify the exposition we will assume in the rest of this paper that $\chi=X$.

The next two results characterize $D\chi\ast$ as a passive convolution operator and these results follow immediately from \cite[Thm 8.12-1]{Zema72}, \cite[Lemma 7.2-1]{Zema72}, the existence of antiderivatives of distributions as discussed in Sec.~\ref{AppdxOnAntiderivativesOfCertainDistributions}, the relationship between the Laplace and Fourier transforms as discussed in Sec.~\ref{AppdxSecNotationsConventionsDefinitions}, and the relationship between positive* mappings \cite[Def.~8.11-1]{Zema72} and Herglotz functions as defined in Sec.~\ref{AppdxSecNotationsConventionsDefinitions} (namely, if $Y:C_+\rightarrow [\mathfrak{H};\mathfrak{H}]$ is a positive* mapping then $h:\mathbb{C}_+\rightarrow [\mathfrak{H};\mathfrak{H}]$ defined by $h(\omega)=iY(-i\omega)$ is a Herglotz function). First, the following three statements are equivalent: i) $D\chi\ast$ is a passive convolution operator on $\mathscr{D}(\mathfrak{H})$; ii) $D\chi$ is Laplace-transformable with a strip of definition $C_+$ and its Laplace-transform $Y=\mathfrak{L}(D\chi):C_+\rightarrow [\mathfrak{H};\mathfrak{H}]$ is a positive* mapping; iii) $D\chi$ has a Fourier transform $\widehat {D\chi}$ in $\mathbb{C}_+$ and $h=i\widehat {D\chi}:\mathbb{C}_+\rightarrow [\mathfrak{H};\mathfrak{H}]$ is a Herglotz function. In particular, the Laplace transform $Y$ in (ii) and the Herglotz function $h$ in (iii) are related by $h(\omega)=iY(-i\omega)$. Second, to every Herglotz function $h:\mathbb{C}_+\rightarrow [\mathfrak{H};\mathfrak{H}]$ there exists a unique distribution $D\chi$ (with $\chi\in [\mathscr{D}(\mathfrak{H});\mathfrak{H}]$) such that it has a Fourier transform $\widehat {D\chi}$ in $\mathbb{C}_+$ with $h=i\widehat{D\chi}$.

Now since $D\chi$ has a Fourier transform $\widehat {D\chi}$ in $\mathbb{C}_+$ and the antiderivative $\chi$ has the property $\operatorname{supp}\chi\subseteq [0,\infty)$ then this implies by statement (ii) in Sec.~\ref{AppdxOnAntiderivativesOfCertainDistributions} and its proof that $\chi$ has a Fourier transform $\widehat {\chi}$ in $\mathbb{C}_+$ and the Herglotz function $h=i\widehat {D\chi}:\mathbb{C}_+\rightarrow [\mathfrak{H};\mathfrak{H}]$ satisfies
\begin{align}\label{AppdxProofOfMainResultsHerglotzFuncFormInSusceptibility}
h(\omega)=i\widehat{D\chi}(\omega)=\omega\widehat{\chi}(\omega),\;\;\;\omega\in\mathbb{C}_+.
\end{align}
This completes the proof of the results in Sec.~\ref{Sec:HomogPassivity}, \ref{Sec:PeriodicPassivity}.

\subsubsection{Proofs of Results in Sec.~\ref{Sec:HomogTranspWnd}, \ref{Sec:PeriodicTranspWnd}, \ref{Sec:Achieving_c}}\label{AppdxSecProofsofSecHomogTranspWndSecPeriodicTranspWndAndSecAchievingc}
In this section we prove the statements from Sec.~\ref{Sec:HomogTranspWnd}, \ref{Sec:PeriodicTranspWnd}, and \ref{Sec:Achieving_c} that were left to be proved here. Our proofs of all these results are essentially based on the connection between the well-known integral representation of a Herglotz function from \cite[Theorem 5.4]{Gesz00}, \cite[Eqs.~(1.1)--(1.3)]{Gesz01}, and \cite[Theorem A.4]{Gesz13} to the Schwindt's representation of a positive* mapping \cite[Theorem 8.11-2]{Zema72} and the characterization of passive convolution operators in terms of positive* mappings \cite[Theorem 8.12-1]{Zema72}. 

More specifically, we first use the integral representation (\ref{AppdxIntegralRepresHerglotzFuncTemperedPOMeasure}), (\ref{AppdxIntegralRepresHerglotzFuncTemperedPOMeasureCoeffFormula}) for the Herglotz function $h(\omega)=\omega \widehat \chi(\omega)$ in (\ref{AppdxProofOfMainResultsHerglotzFuncFormInSusceptibility}) in terms of an operator-valued measure $\Omega$ that quantifies losses via the identity (\ref{AppdxIntegralRepresHerglotzFuncTemperedPOMeasure&Connection2Losses}). In particular, this identity proves the equality between (\ref{HomoImhGenConn2Passivity0}) and (\ref{HomoImhGenConn2Passivity}). Next, we prove the formula (\ref{AppdxIntegralRepresHerglotzFuncTemperedPOMeasureIntervalMeasFormula}) which gives the correspondence between losses, $\Im\,h(\omega)$, and the measure $\Omega$ and hence proves the identity (\ref{HomogMeasureLosses}). Finally, we prove the five main consequences (labeled as (i)--(v) in the subsection below entitled, ``On transparency windows'') of a transparency window defined in (\ref{DefTranspWindHomog}) as a frequency interval $(\omega_1,\omega_2)\subseteq\mathbb{R}$ in which $\Omega((\omega_1,\omega_2))=0$. In particular, using the integral representation (\ref{AppdxIntegralRepresHerglotzFuncTemperedPOMeasure}), (\ref{AppdxIntegralRepresHerglotzFuncTemperedPOMeasureCoeffFormula}) for $h(\omega)$ in (\ref{AppdxProofOfMainResultsHerglotzFuncFormInSusceptibility}) we prove the main consequences of a transparency window $(\omega_1,\omega_2)$, namely, analytic continuation of $h(\omega)$ across $(\omega_1,\omega_2)$ to a bounded-operator-valued function satisfying the self-adjoint and monotonicity conditions (\ref{HerglotzHomogMonot}). And, moreover, we also prove the fact mentioned in Sec.~\ref{Sec:Achieving_c} on achieving the upper bound $||{\bf v}_e||=c$ that the condition (\ref{HerglotzPlanewaveNullspace}) being true for $h$ at a single frequency in a transparency window implies that (\ref{HerglotzPlanewaveNullspace}) and (\ref{HerglotzPlanewaveNullspace1}) are true for all frequencies.

\paragraph*{Integral representations}
As we proved in Sec.~\ref{AppdxSecProofsofSecHomogPassivityAndSecPeriodicPassivity}, $h=i\widehat {D\chi}:\mathbb{C}_+\rightarrow [\mathfrak{H};\mathfrak{H}]$ is a Herglotz function. Hence, it follows from the well-known results on the integral representation of a Herglotz function (see, for instance, \cite{Aron56}, \cite{Gesz00}, \cite{Gesz01}, \cite[Appendix A]{Gesz13}, and references within) that there exists a unique tempered PO measure $\Omega:\mathfrak{B}_{\mathbb{R},\infty}\rightarrow [\mathfrak{H};\mathfrak{H}]_+$ (see \cite[Def.~8.7-1]{Zema72}, where $\mathfrak{B}_{\mathbb{R},\infty}$ denotes the bounded Borel subsets of $\mathbb{R}$) such that 
\begin{align}
g(\lambda)=(1+\lambda^2)^{-1}
\end{align}
is integrable with respect to the measure $\Omega$ and
\begin{align}
h(\omega)=h_0+ h_1\omega+\int_{\mathbb{R}}d{\Omega_\lambda}\left(\frac{1}{\lambda-\omega}-\frac{\lambda}{1+\lambda^2}\right)\label{AppdxIntegralRepresHerglotzFuncTemperedPOMeasure}
\end{align}
for every $\omega\in \mathbb{C}_+$, where
\begin{align}
h_0=\Re\,h(i),\;\;\;0\leq h_1=s\mbox{-}\lim_{\eta\uparrow\infty}\frac{1}{i\eta}h(i\eta),\label{AppdxIntegralRepresHerglotzFuncTemperedPOMeasureCoeffFormula}
\end{align}
where $\Re\,h(i)$ denotes the real part of the operator $h(i)$ and $s\mbox{-}\lim_{\eta\uparrow\infty}$ means the limit in the strong operator topology on $[\mathfrak{H};\mathfrak{H}]$ as real $\eta\rightarrow +\infty$. And, moreover, the Stieltjes inversion formula for the measure $\Omega$ holds in the strong operator topology, i.e.,
\begin{align}\label{AppdxStieltjesInversionFormula}
\frac{1}{2}\Omega((\omega_1,\omega_2))+\frac{1}{2}\Omega([\omega_1,\omega_2])=s\mbox{-}\lim_{\eta\downarrow 0}\int_{\omega_1}^{\omega_2}\frac{1}{\pi}\Im\,h(\lambda+i\eta)d\lambda
\end{align}
for any $\omega_1,\omega_2\in \mathbb{R}$ with $\omega_1<\omega_2$, where the integral in (\ref{AppdxStieltjesInversionFormula}) is the Bochner integral (see \cite[Appendix G]{Zema72} for definition) of the function $\lambda\mapsto \Im\,h(\lambda+i\eta)$ with respect to the Lebesgue measure on $\mathfrak{B}_\mathbb{R}$ (the Borel subsets of $\mathbb{R}$) and $s\mbox{-}\lim_{\eta\downarrow 0}$ means the limit in the strong operator topology on $[\mathfrak{H};\mathfrak{H}]$ as positive real $\eta\rightarrow 0$. Hence, from formula (\ref{AppdxStieltjesInversionFormula}) and \cite[Lemma 2.2-1]{Zema72} it follows that 
\begin{align}
\Omega((\omega_1,\omega_2))=s\mbox{-}\lim_{\delta\downarrow 0}s\mbox{-}\lim_{\eta\downarrow 0}\int_{\omega_1+\delta}^{\omega_2-\delta}\frac{1}{\pi}\Im\,h(\lambda+i\eta)d\lambda\label{AppdxIntegralRepresHerglotzFuncTemperedPOMeasureIntervalMeasFormula}
\end{align}
for any $\omega_1,\omega_2\in \mathbb{R}$ with $\omega_1<\omega_2$.

We will now prove that the following equality holds
\begin{align}
\Re\,\int_{-\infty}^{\infty}(v(t),D\chi\ast v(t))dt=\frac{1}{2}\int_{\mathbb{R}}d(\Omega_\omega \widehat v(\omega),\widehat v(\omega))\label{AppdxIntegralRepresHerglotzFuncTemperedPOMeasure&Connection2Losses}
\end{align}
for every $v\in \mathscr{D}(\mathfrak{H})$, where the integral on the right in (\ref{AppdxIntegralRepresHerglotzFuncTemperedPOMeasure&Connection2Losses}) is defined as in \cite[\S 2.6 \& \S 8.7]{Zema72} for the Fourier transform defined by (\ref{AppdxDefFourierTransformTestingFuncRapidDescent}). In fact, it follows from \cite[Thm 8.7-3]{Zema72} and the polarization identity for sesquilinear forms \cite[Appendix A6]{Zema72}, that the tempered PO measure $\Omega$ is uniquely determined by the positive quadratic form $v\mapsto\Re\,\int_{-\infty}^{\infty}(v(t),D\chi\ast v(t))dt$ on $\mathscr{D}(\mathfrak{H})$.

To prove (\ref{AppdxIntegralRepresHerglotzFuncTemperedPOMeasure&Connection2Losses}), we begin by using \cite[Theorem 2.3-3]{Zema72} to define a PO measure $Q:\mathfrak{B}_{\mathbb{R}}\rightarrow [\mathfrak{H};\mathfrak{H}]_+$ (where $\mathfrak{B}_{\mathbb{R}}$ denotes the Borel subsets of $\mathbb{R}$) by
\begin{align}
Q(B)=\int_{B}d\Omega_\lambda g(\lambda),\;\;\;B\in \mathfrak{B}_{\mathbb{R}}.\label{AppdxDefPOMeasForHerglotzFuncInTermsOfTheTemperedMeas}
\end{align}
Thus, since
\begin{align}
\frac{1+\lambda\omega}{\lambda-\omega}=g(\lambda)^{-1}\left(\frac{1}{\lambda-\omega}-\frac{\lambda}{1+\lambda^2}\right),\label{AppdxIdentityIntegrandInIntegralRepHerglotzFunc}
\end{align}
it follows from \cite[Thm 2.3-4]{Zema72} that
\begin{align}
h(\omega)=h_0+ h_1\omega+\int_{\mathbb{R}}d{Q_\lambda}\frac{1+\lambda\omega}{\lambda-\omega}
\end{align}
for every $\omega\in \mathbb{C}_+$.

Next, using the change-of-variable $\zeta=-i\omega$, it follows that the function $Y:C_+\rightarrow [\mathfrak{H};\mathfrak{H}]$ defined by $Y(\zeta)=-ih(i\zeta)$ is a positive* mapping and is the Laplace transform $Y=\mathfrak{L}(D\chi):C_+\rightarrow [\mathfrak{H};\mathfrak{H}]$ with the Schwindt's representation (cf.~\cite[Thm 8.11-2]{Zema72})
\begin{align}
Y(\zeta)=P_0 +P_1\zeta+\int_{\mathbb{R}}dP_\eta\frac{1-i\eta\zeta}{\zeta-i\eta},
\end{align}
where $P_0, P_1$, and the PO measure $P:\mathfrak{B}_{\mathbb{R}}\rightarrow [\mathfrak{H};\mathfrak{H}]_+$ are related to $h_0,h_1$ and $Q$ by the formulas
\begin{align}
P_0=-ih_0,\;\;\;P_1=h_1,\;\;\;dP_\eta=dQ_{-\eta},
\end{align} 
i.e., $P(B)=Q(-B)$ for every $B\in \mathfrak{B}_{\mathbb{R}}$,
and, in particular, $\int_{\mathbb{R}}dP_{\eta}f(\eta)=\int_{\mathbb{R}}dQ_{\eta}f(-\eta)$ for every $f\in \mathscr{G}_{\mathbb{R}}$ (where $\mathscr{G}_{\mathbb{R}}$ denotes the set of all bounded Borel functions on $\mathbb{R}$). Finally, the proof of the equality (\ref{AppdxIntegralRepresHerglotzFuncTemperedPOMeasure&Connection2Losses})  now follows immediately from these facts and the proof of \cite[Thm 8.12-1]{Zema72}.

\paragraph*{On transparency windows}
In the rest of this section we will assume that $(\omega_1,\omega_2)\subseteq\mathbb{R}$, $\omega_1<\omega_2$ is a transparency window for the medium with susceptibility $\chi$, i.e.,
\begin{align}
\Omega((\omega_1,\omega_2))=0.\label{AppdxProofTransparencyWindow}
\end{align}
From the linear systems theory perspective, the condition (\ref{AppdxProofTransparencyWindow}) interpreted in terms of the equality (\ref{AppdxIntegralRepresHerglotzFuncTemperedPOMeasure&Connection2Losses}) just means the linear system characterized by the passive convolution operator $D\chi\ast$ on $\mathscr{D}(\mathfrak{H})$ is lossless in the interval $(\omega_1,\omega_2)$.

Now we denote below the complement of the set $(\omega_1,\omega_2)$ in $\mathbb{R}$ by $E$, i.e.,
\begin{align}
E=(-\infty, \omega_1]\cup [\omega_2,\infty).
\end{align}
Then condition (\ref{AppdxProofTransparencyWindow}) implies the Herglotz function $h=i\widehat {D\chi}:\mathbb{C}_+\rightarrow [\mathfrak{H};\mathfrak{H}]$ has the following properties (i)--(v) which we first state below and then prove after.

(i) First, the domain of definition of $h$ can be extended to $\mathbb{C}\setminus E$ by defining it in terms of the integral
\begin{align}
h(\omega)=h_0+ h_1\omega+\int_{E}d{\Omega_\lambda}\left(\frac{1}{\lambda-\omega}-\frac{\lambda}{1+\lambda^2}\right)\label{AppdxTranspWindIntegralRepresHerglotzFuncTemperedPOMeasure}
\end{align}
for every $\omega\in\mathbb{C}\setminus E$.

(ii) Next, the function $h:\mathbb{C}\setminus E\rightarrow [\mathfrak{H};\mathfrak{H}]$ in (\ref{AppdxTranspWindIntegralRepresHerglotzFuncTemperedPOMeasure}) is an analytic $[\mathfrak{H};\mathfrak{H}]$-valued function in its domain. In particular, this function is just the analytical continuation of the Herglotz function $h:\mathbb{C}_+\rightarrow [\mathfrak{H};\mathfrak{H}]$ into the open lower-half of the complex plane through the interval $(\omega_1,\omega_2)$. 

(iii) Next, differentiation under the integral in (\ref{AppdxTranspWindIntegralRepresHerglotzFuncTemperedPOMeasure}) is allowed and the derivative of $h$ has the integral representation
\begin{align}
h^{\prime}(\omega)=h_1+\int_{E}d{\Omega_\lambda}\frac{1}{(\lambda-\omega)^2}\label{AppdxDerivativeTranspWindIntegralRepresHerglotzFuncTemperedPOMeasure}
\end{align}
for every $\omega\in\mathbb{C}\setminus E$.

(iv) Next, the function $h$ is self-adjoint and monotonic in $(\omega_1,\omega_2)$, i.e.,
\begin{align}
\Im\,h(\omega)=0,\;h^{\prime}(\omega)\geq 0,\;\;\;\omega\in (\omega_1,\omega_2).
\end{align}

(v) Finally, if $\psi\in \mathfrak{H}$ and $h^{\prime}(\omega_0)\psi=0$ for some $\omega_0\in (\omega_1,\omega_2)$ then
\begin{align}
h(\omega)\psi=h_0\psi,\;\;\;\Im\,h_0=0
\end{align}
for every $\omega\in \mathbb{C}\setminus E$. Hence, extending this identity by continuity to $E$ implies $h(\omega)\psi=h_0\psi$ for all $\omega\in \mathbb{C}$.

We will now prove these statements (i)--(v). We begin with the proof of statement (i). First, it follows immediately from the assumption (\ref{AppdxProofTransparencyWindow}), the identity (\ref{AppdxIdentityIntegrandInIntegralRepHerglotzFunc}), 
and the elementary properties of $\sigma$-finite PO measures (namely, \cite[Thm 2.3-4]{Zema72}, \cite[Thm 2.2-1]{Zema72}, and the identity \cite[p.~25, Eq.~(4)]{Zema72}) that for every $\omega\in \mathbb{C}_+$ we have
\begin{align}
\int_{\mathbb{R}}d\Omega_\lambda\left(\frac{1}{\lambda-\omega}-\frac{\lambda}{1+\lambda^2}\right)&=\int_{E}d\Omega_\lambda\left(\frac{1}{\lambda-\omega}-\frac{\lambda}{1+\lambda^2}\right),
\end{align}
where the integral on the right is with respect to the restriction of the measure $\Omega$ to $\mathfrak{B}_{E,\infty}$ (where $\mathfrak{B}_{E,\infty}$ denotes the bounded Borel subsets of $E$). Furthermore, by the identity (\ref{AppdxIdentityIntegrandInIntegralRepHerglotzFunc}) and \cite[Thm 2.3-4]{Zema72} it follows that
\begin{align}
\int_{E}d\Omega_\lambda\left(\frac{1}{\lambda-\omega}-\frac{\lambda}{1+\lambda^2}\right)
=\int_{E}d{Q_\lambda}\frac{1+\lambda\omega}{\lambda-\omega},\label{AppdxEqOfIntegralsInTranspWndInIntReprOfHerglotzFunc}
\end{align}
where $Q:\mathfrak{B}_{E}\rightarrow[\mathfrak{H};\mathfrak{H}]_+$ is the PO measure defined above in (\ref{AppdxDefPOMeasForHerglotzFuncInTermsOfTheTemperedMeas}) restricted to $\mathfrak{B}_{E}$ (where $\mathfrak{B}_{E}$ denotes the Borel subsets of $E$). Moreover, it follows from the identity (\ref{AppdxIdentityIntegrandInIntegralRepHerglotzFunc}) and \cite[Thm 2.3-4]{Zema72} that the integrals in (\ref{AppdxEqOfIntegralsInTranspWndInIntReprOfHerglotzFunc}) are well-defined and equal for all $\omega\in \mathbb{C}\setminus E$. From these facts, the proof of statement (i) now follows immediately.

We will now prove statements (ii) and (iii). Consider the function
\begin{align}
f(\omega)(\lambda)=\frac{1+\lambda\omega}{\lambda-\omega},\;\;\;\omega\in \mathbb{C}\setminus E,\;\lambda\in E.
\end{align}
Then $f(\omega)\in \mathscr{G}_E$ for every $\omega\in \mathbb{C}\setminus E$ (where $\mathscr{G}_E$ denotes the set of all bounded Borel functions on $E$) and the $\mathscr{G}_E$-valued function $f:\mathbb{C}\setminus E\rightarrow \mathscr{G}_E$
is analytic on the open set $\mathbb{C}\setminus E$ into the Banach space $\mathscr{G}_E$ with the sup norm $||\cdot||_{\infty}$. Also, the integral map $\mathcal{I}_Q:\mathscr{G}_E\rightarrow [\mathfrak{H};\mathfrak{H}]$ on $\mathscr{G}_E$ defined by
\begin{align}
\mathcal{I}_Q(u)=\int_EdQ_{\lambda}u(\lambda),\;\;\;u\in \mathscr{G}_E
\end{align} 
is a continuous linear operator which follows from \cite[Thm 2.2-3]{Zema72} and the inequality \cite[p.~25, Ineq.~(2)]{Zema72}. This implies the composition of the two maps $\mathcal{I}_Q\circ f:\mathbb{C}\setminus E\rightarrow [\mathfrak{H};\mathfrak{H}]$ is an analytic $[\mathfrak{H};\mathfrak{H}]$-valued function which satisfies the equality $\frac{d}{d\omega}(\mathcal{I}_Q\circ f)=\mathcal{I}_Q\circ \frac{df}{d\omega}$ on $\mathbb{C}\setminus E$, where  
\begin{align}
\frac{df}{d\omega}(\omega)(\lambda)=\frac{g(\lambda)^{-1}}{(\lambda-\omega)^2},\;\;\;\omega\in \mathbb{C}\setminus E,\;\lambda\in E.
\end{align}
It follows now immediately from these facts, \cite[Thm 2.3-4]{Zema72}, and the equality of the integrals in (\ref{AppdxEqOfIntegralsInTranspWndInIntReprOfHerglotzFunc}) that
\begin{align}
\frac{d}{d\omega}\int_{E}d\Omega_\lambda\left(\frac{1}{\lambda-\omega}-\frac{\lambda}{1+\lambda^2}\right)
=\int_{E}d\Omega_\lambda\frac{1}{(\lambda-\omega)^2}\label{AppdxDiffUnderTheIntegralSign}
\end{align}
for $\omega\in\mathbb{C}\setminus E$, where $\frac{d}{d\omega}$ here denotes the derivative in the operator norm topology of $[\mathfrak{H};\mathfrak{H}]$.  From these facts and statement (i), the proof of statements (ii) and (iii) now follow immediately.

We will now prove statements (iv) and (v). First, it follows from the identity \cite[p.~25, Eq.~(4)]{Zema72}, \cite[Thm 2.2-1]{Zema72}, and (\ref{AppdxEqOfIntegralsInTranspWndInIntReprOfHerglotzFunc}) that
\begin{gather}
\left(\psi,\int_{E}d\Omega_{\lambda}\left(\frac{1}{\lambda-\omega}-\frac{\lambda}{1+\lambda^2}\right)
\psi\right)=\int_{E}d(Q_{\lambda}\psi,\psi)\frac{1+\lambda\omega}{\lambda-\omega}\label{AppdxTranspWindInnerProdIntegralRepresHerglotzFuncTemperedPOMeasure}\\
\left(\psi,\int_{E}d\Omega_{\lambda}\frac{1}{(\lambda-\omega)^2}\psi\right)=\int_{E}d(Q_{\lambda}\psi,\psi)\frac{g(\lambda)^{-1}}{(\lambda-\omega)^2}\nonumber
\end{gather}
for any $\psi\in \mathfrak{H}$ and any $\omega\in\mathbb{C}\setminus E$, where the integrals in (\ref{AppdxTranspWindInnerProdIntegralRepresHerglotzFuncTemperedPOMeasure}) on the right are Lebesgue integrals with respect to the positive finite measure $(Q(\cdot)\psi,\psi):\mathfrak{B}_E\rightarrow [0,\infty)$. Now the statement (iv), (v) immediately follows from the representation (\ref{AppdxTranspWindInnerProdIntegralRepresHerglotzFuncTemperedPOMeasure}), the fact that for all $\omega\in (\omega_1,\omega_2)$ and $\lambda\in E$ the integrands in (\ref{AppdxTranspWindInnerProdIntegralRepresHerglotzFuncTemperedPOMeasure}) are real with $\frac{g(\lambda)^{-1}}{(\lambda-\omega)^2}>0$, the fact that $h_0=\Re\,h(i)$ is self-adjoint, $h_1\geq 0$, and from statements (i), (iii). This completes the proof of statements (i)--(v). These statements together with the equality (\ref{AppdxProofOfMainResultsHerglotzFuncFormInSusceptibility}) prove our results in Sec.~\ref{Sec:HomogTranspWnd}, \ref{Sec:PeriodicTranspWnd}, \ref{Sec:Achieving_c}.

\bibliography{SpeedLightLimitations}

\end{document}